\documentclass{ws-ijmpd}

\begin{document}

\markboth{Authors' Names}
{Instructions for Typing Manuscripts (Paper's Title)}

%
\catchline{}{}{}{}{}
%

\title{WARPED SELF-GRAVITATING U(1) GAUGE COSMIC STRINGS IN 5D  }

\author{REINOUD JAN SLAGTER}

\address{Dept of Physics, Univ of Amsterdam and ASFYON \\
Astronomisch Fysisch Onderzoek Nederland, 1405EP Bussum, The Netherlands\\
info@asfyon.nl}

\author{DERK MASSELINK}

\address{ASFYON, The Netherlands\\
info@asfyon.nl}

\maketitle

\begin{history}
\received{Day Month Year}
\revised{Day Month Year}
\comby{Managing Editor}
\end{history}

\markboth{REINOUD JAN SLAGTER and DERK MASSELINK}
{WARPED SELF-GRAVITATING U(1) GAUGE COSMIC STRINGS IN 5D }

\begin{abstract}

We present the "classical" Nielsen-Olesen vortex solution on a warped 5-dimensional spacetime, where we solved the effective
4-dimensional equations from the 5-dimensional equations together with the junction and boundary conditions.
4-dimensional cosmic strings show some serious problems concerning the mechanism of string smoothing related to the string mass per unit length, $G\mu \leq 10^{-6}$. Moreover,  there is no observational evidence of axially symmetric lensing effect caused by cosmic strings.
Also super-massive cosmic strings ($G\mu\gtrsim 1$), predicted by superstring theory, possess some problems. They are studied because the universe may have undergone phase transitions at scales much higher than the GUT scale. But $G\mu \gtrsim 1$ is far above observational bounds, so one needs  an inflationary scenario to smooth them out. Further, it is believed that
these super-massive strings never extended to macroscopic size.
Brane world models could overcome these problems. $G\mu$ could be warped down to GUT scale, even if its value was at the Planck scale.

In our warped cosmic string model, where the string  mass  per unit length in the bulk can be of order of the Planck scale, we find that  the  4-dimensional brane spacetime is exponential warped down. Moreover, asymptotically the induced 4-dimensional spacetime does not show conical behavior. So there is no angle deficit compared to its value in the bulk and the spacetime seems to be unphysical, at least under fairly weak assumptions on the stress-energy tensor and without a positive brane tension.
The results are confirmed by numerical solutions of the field equations.

\end{abstract}


\section{Introduction}

The standard model is extremely successful up to scales $M_{EW}\sim 10^3$ GeV. The fundamental scale of gravity is the Planck scale $M_{Pl}\sim 10^{19}$ GeV. It is the scale where quantum gravity will act. The discrepancy between these two scales is called the hierarchy problem. Electro-weak interactions have been tested up to $M_{EW}$, while gravity,
on the other hand, has been tested to several millimeters, 32 orders of magnitude above $M_{pl}$.
Brane world models could overcome the hierarchy problem. The idea originates from string theory.
One of the predictions of string theory is the existence of branes embedded in the full bulk space time. Gravitons can then propagate into the bulk, while other fields are confined to these branes. It also predicts that space time is 10-dimensional, with 6 of them are very compact and small, not verifiable by any experiment. There are many models which attacked the hierarchy problem. Essentially there are globally two categories: flat compact extra dimensions\cite{ark} and warped extra dimensions\cite{Shir}.
Recently, there is growing interest in the second category, i.e., the Randall-Sundrum(RS) warped 5-dimensional geometry\cite{Ran1,Ran2}. We live in a 3+1 dimensional space time embedded in a 5-dimensional space time, with an extra dimension which can be very large compared to the ones predicted in string theory. One estimates that the extra dimension can be as large as $10^{-3} cm$, which is the under-bound of Newton's law in our world.
The observed 4-dimensional Planck scale $M_{Pl}\equiv M_4$ is no longer the fundamental scale but an effective one, an important consequence of the extra dimensions, which is now  $M_5$, the Planck scale in 5D. If we consider the Einstein equations in 5D,
\begin{equation}
{^{(5)}}G_{\mu\nu}=-\Lambda_5 {^{(5)}}g_{\mu\nu}+\kappa_5^2 \Bigl({^{(5)}}T_{\mu\nu}+T^{brane}_{\mu\nu}\Bigr),\label{eqn1}
\end{equation}
with $T^{brane}_{\mu\nu}={^{(4)}T}_{\mu\nu}-\lambda_4 {^{(4)}g}_{\mu\nu}$ and y the extra dimension, then  $\kappa_5^2 =8\pi G_5=\frac{8\pi}{M_5^3}$ is the gravitational coupling constant. If $L$ is the length scale of the extra dimension, then
$M^2_{Pl}\sim M^3_5 L$. So if the extra-dimensional volume is the Planck scale, i.e., $L\sim \frac{1}{M_{Pl}}$, then $M_5\sim M_4$.
But if the extra-dimensional volume is significantly larger then the Planck scale, then the true fundamental scale $M_5$ can be much smaller
then the effective scale $M_4\sim 10^{19}GeV$. So the weakness of gravity can be understand by the fact that it "spreads" into the extra dimension
and only a part is felt in 4D. If $L$ is of order $10^{-1} mm \sim (10^{-15} TeV)^{-1}$, then  $M_5 \sim 10^9 GeV$, much smaller than the observed $10^{19} GeV$.
The RS brane world models will further lower down the $M_5$ scale, by considering warped space times.
In this article we follow the formulation and notation of the brane-world gravity models of Maartens\cite{Roy} and Durrer\cite{Dur}

In this paper we will investigate on an axially symmetric 5-dimensional warped space time the modifications of the behavior of a gauge cosmic string in the Abelian Higgs model.
Cosmic strings occur as topological
defects, consisting of confined regions of false vacuum in gauge theories with spontaneous symmetry breaking. If local strings appeared in phase transitions in the early universe,
they could have served as seeds for the formation of galaxies. However, observations of the cosmic microwave background, would rule out this model. M-theory, the improved
version of superstring theory, allows, via brane-world scenarios, macroscopic fundamental strings that could play a role very similar to that of cosmic strings.
The resulting super-massive cosmic strings are even more exotic, because they could develop  singular behavior at finite distance of the core of the string.

In section 2 we outline the model under consideration and present some numerical solutions. In section 3 we derive the field equations on the brane. In section 4 we investigate
the angle deficit and the changes with respect to the 4D model.
In the appendix A we give a brief overview of the 4D Nielsen-Olesen U(1) gauge cosmic string and his features. We used the Grtensor program in Maple 13 to check the equations.

\section{The Model}

Let us consider the 5D model\cite{Aoy}
\begin{eqnarray}
{\cal S}=\int d^5x\sqrt{-^{(5)}g}\Bigl[\frac{1}{2\kappa_5^2}({^{(5)}R}-\Lambda_5)+S_{bulk}\Bigr]
+\int d^4x\sqrt{-^{(4)}g}\Bigl[\frac{1}{\kappa_5^2}\lambda_4+  S_{brane} \Bigr],\label{eqn2}
\end{eqnarray}
with $\Lambda_5$ the cosmological constant in the bulk, $\lambda_4$ the brane tension, $S_{bulk}$ the matter Lagrangian in the bulk and $S_{brane}$ the effective 4D Lagrangian, which is given by a generic functional of the brane metric and matter fields on the brane and will also contain the extrinsic curvature corrections due to the projection of the 5D curvature.
If there is a bulk scalar field ${^{(5)}\Phi}$ (no coupling to a 5D gauge field $A_\mu$ , as in the 4D case), which could stabilize the branes\cite{gold}, then we have for
the 5D equations (from now on all the indices run from 0..4)
\begin{eqnarray}
{^{(5)}G}_{\mu\nu}=-\Lambda_5{^{(5)}g}_{\mu\nu}+\kappa_5^2{^{(5)}T}_{\mu\nu},\quad {^{(5)}\nabla}_\mu {^{(5)}\nabla}^\mu {^{(5)}\Phi}=0,\label{eqn3}
\end{eqnarray}
with ${^{(5)}T}_{\mu\nu}$ the energy momentum tensor of the bulk scalar field and where  we write ${^{(5)}\Phi}={^{(5)}X} e^{i\varphi}$. The equations can easily be obtained from the 4D case in the appendix A, with ${^{(5)}P}=1, {^{(5)}V(^{(5)}\Phi)}=0$.
We will consider here the "classical" Nielsen-Olesen string on a warped 5 dimensional space time
\begin{equation}
ds^2=F(y)\Bigl[e^{A(r,t)}(-dt^2+dz^2)+dr^2+K(r,t)^2e^{-2A(r,t)}d\varphi^2\Bigr]+dy^2.\label{eqn4}
\end{equation}
The 4 dimensional coupled field equations of the U(1)-gauge cosmic string is described by Laguna-Castillo and Matzner\cite{Lag} and Garfinkle \cite{Garf}. See appendix A for an overview.

It turns out that on the space time (4)( for P=0) the y-dependent part is separable. We then obtain for $F(y)$ and ${^{(5)}X}(y)$ the set equations
\begin{equation}
\partial_{yy}F=-\frac{2}{3}\Lambda_5 F-\frac{1}{3}\kappa_5^2F(\partial_y {^{(5)}X})^2+c_1,\label{eqn5}
\end{equation}
\begin{equation}
(\partial_y F)^2=-\frac{2}{3}\Lambda_5 F^2+\frac{1}{3}\kappa_5^2 F^2(\partial_y {^{(5)}X})^2+2c_1F,\label{eqn6}
\end{equation}
\begin{equation}
\partial_{yy} {^{(5)}X}=-2\frac{\partial_y {^{(5)}X}\partial_y F}{F},\label{eqn7}
\end{equation}
with $c_1$ some constant.
For $c_1=0$ we obtain a solution of ${^{(5)}X}$ and $F$ of the form
\begin{equation}
{^{(5)}X}(y)=\frac{\sqrt{3}}{2\kappa_5}\ln\Bigl[\frac{e^{-\frac{2}{3}\sqrt{-6\Lambda_5}y}+\frac{(16\sqrt{3}\kappa_5-4C_2C_3)\Lambda_5}{C_2\sqrt{-6\Lambda_5}}}{e^{-\frac{2}{3}\sqrt{-6\Lambda_5}y}
-\frac{(16\sqrt{3}\kappa_5+4C_2C_3)\Lambda_5}{C_2\sqrt{-6\Lambda_5}}}\Bigr]+C_4,\label{eqn8}
\end{equation}
\begin{equation}
F(y)=\frac{C_1}{\sqrt{\partial_y {^{(5)}X}}},\label{eqn9}
\end{equation}
with $C_i$ some constants. The shape of these solutions depends on the several constants, determined by the junctions conditions.
In figure 1 we represent a typical solution of ${^{(5)}X}$ and F.

\begin{figure}[pb]
\centerline{
\includegraphics[width=5cm]{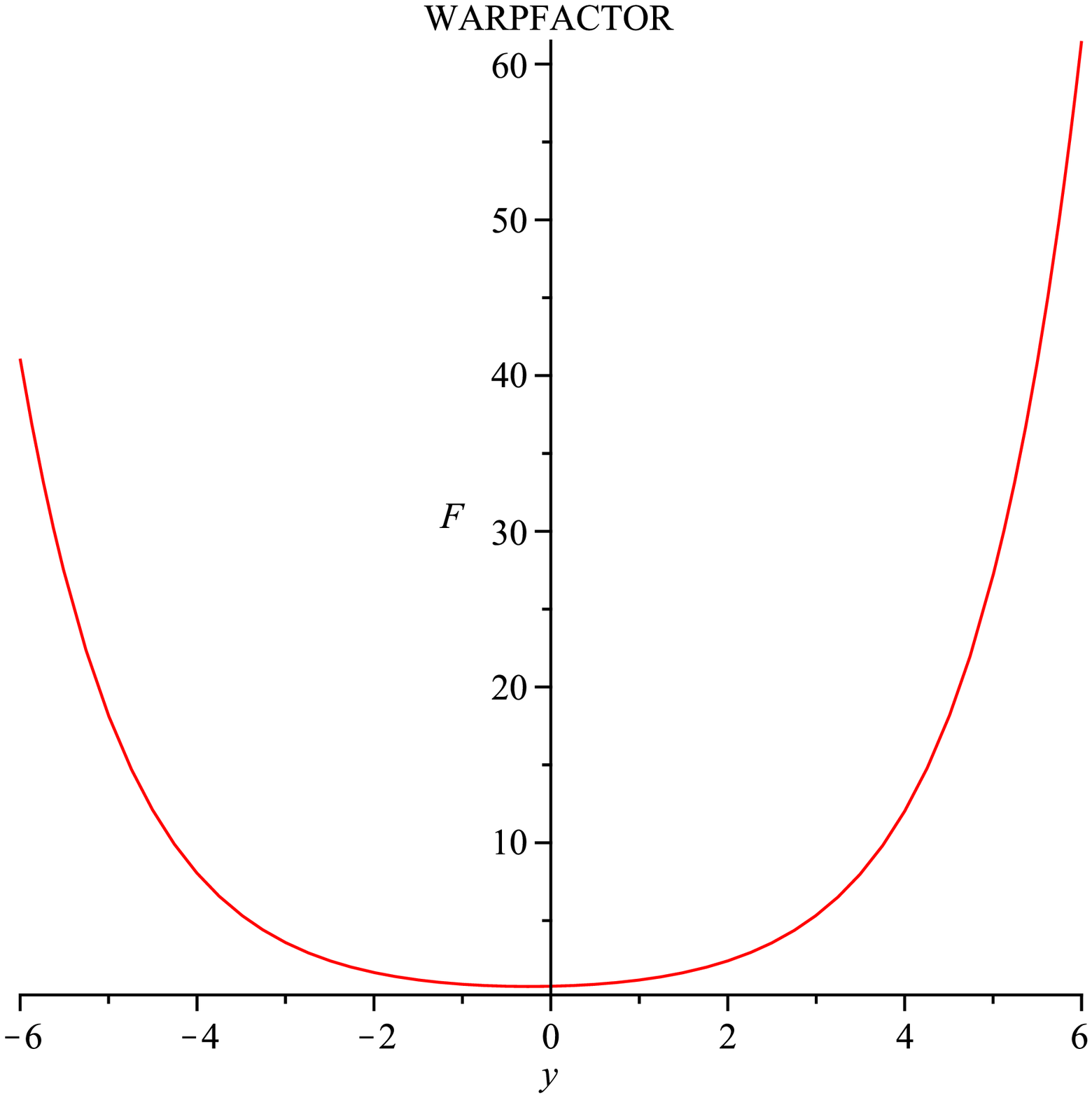}
\includegraphics[width=5cm]{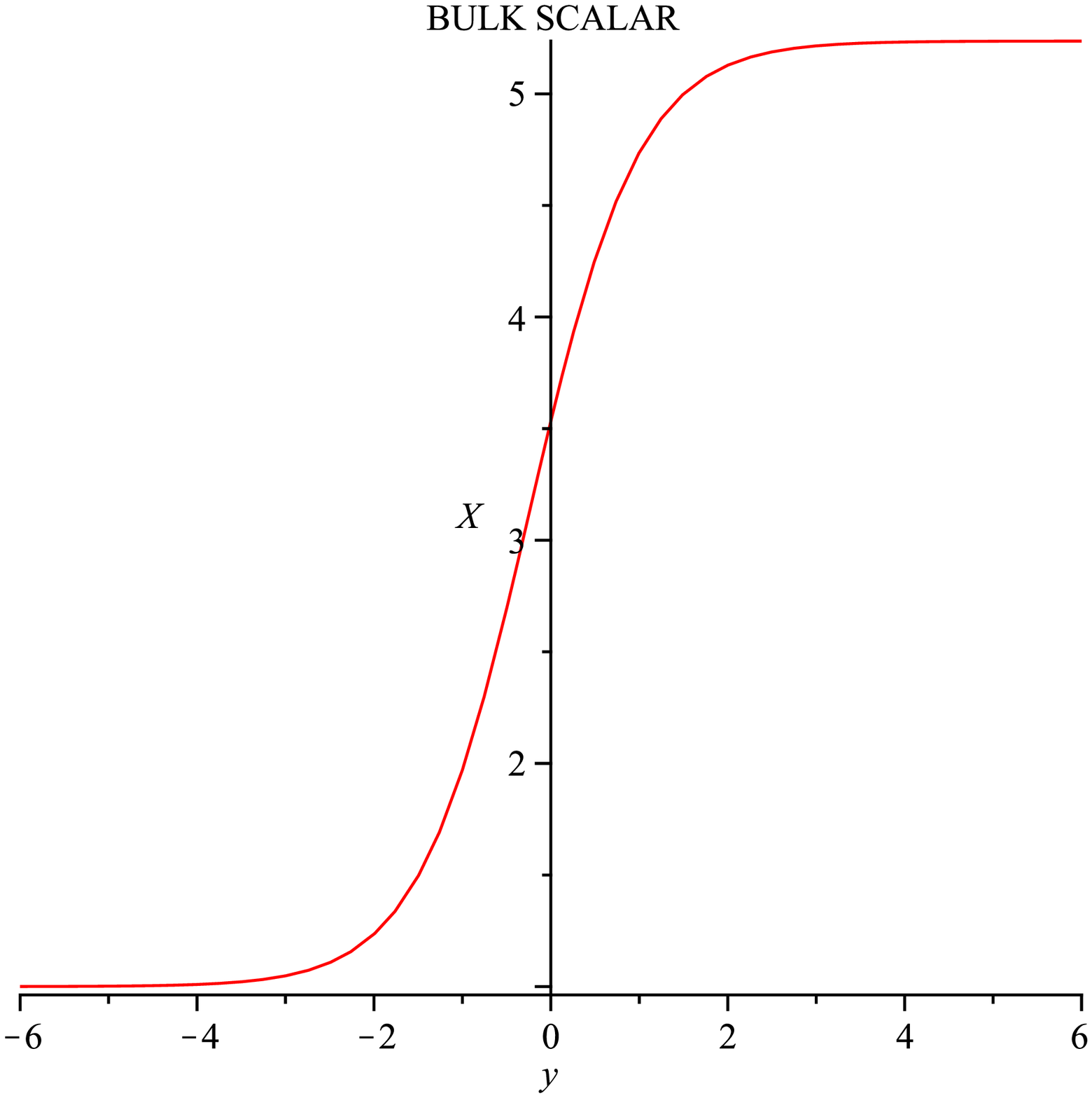}}
\vspace*{8pt}
\caption{Typical solution of the bulk scalar field ${^{(5)}X}$ and warp factor F.\label{f1}}
\end{figure}
It is remarkable that these equations for $F(y)$ and ${^{(5)}X(y)}$ are separable.

From now on we will consider the case of an empty bulk ( $c_1\neq 0$) and the field equations on the brane time-independent. There are 2 solutions for $F$:
\begin{eqnarray}
F(y)=\frac{3c_1}{2\Lambda_5}+D_1\sinh(\frac{1}{3}\sqrt{-6\Lambda_5}y)\pm\frac{\sqrt{9c_1^2+4D_1^2\Lambda_5^2}}{2\Lambda_5}\cosh(\frac{1}{3}\sqrt{-6\Lambda_5}y),\cr
F(y)=\frac{3c_1}{2\Lambda_5}+D_2\cos(\frac{1}{3}\sqrt{6\Lambda_5}y)\pm\frac{\sqrt{9c_1^2-4D_2^2\Lambda_5^2}}{2\Lambda_5}\sin(\frac{1}{3}\sqrt{6\Lambda_5}y),\label{eqn10}
\end{eqnarray}
with $D_i$ some constants.
Typical plots of the two solutions are depicted  in figure 2.

\begin{figure}[pb]
\centerline{
\includegraphics[width=5cm]{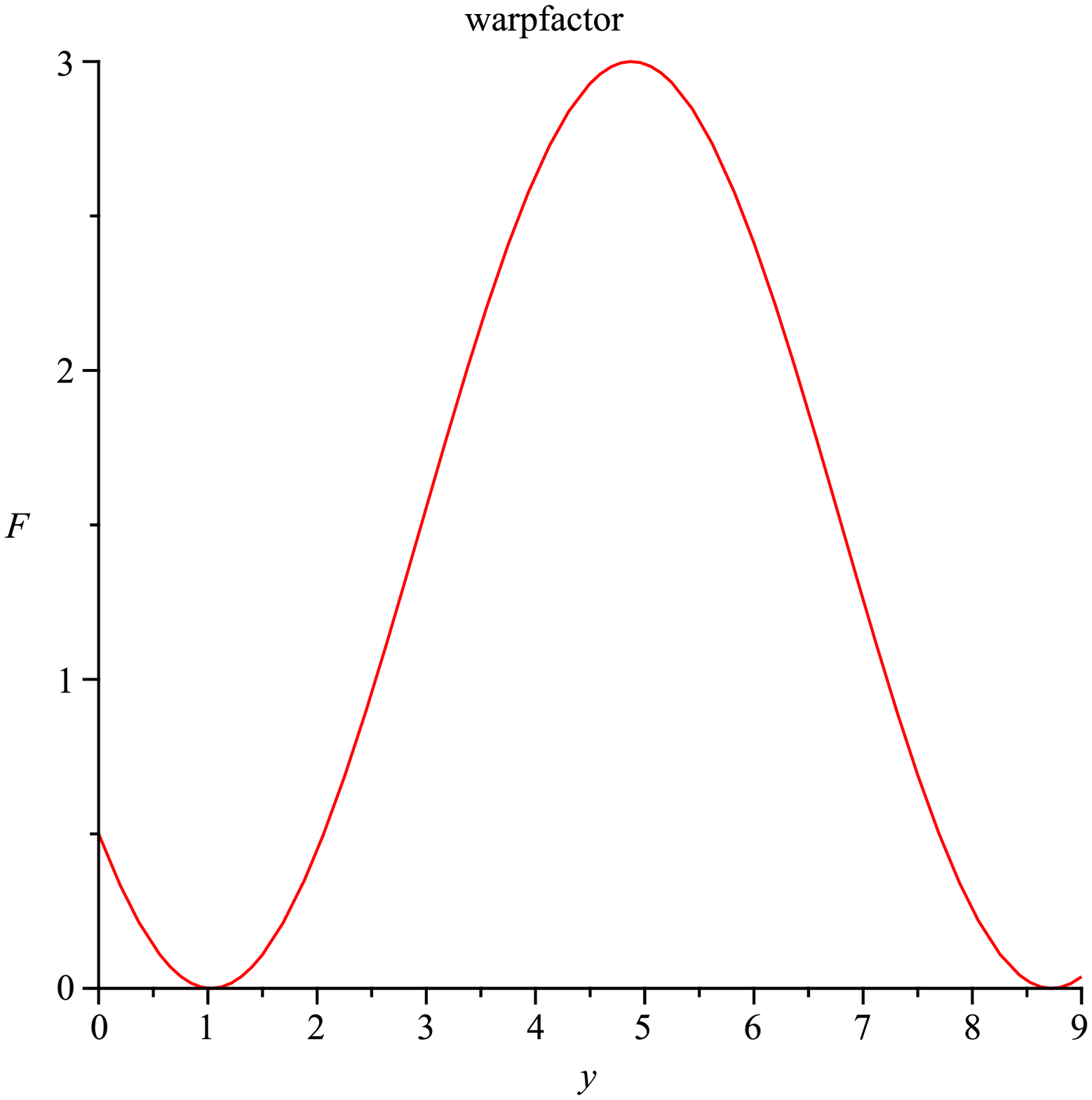}
\includegraphics[width=5cm]{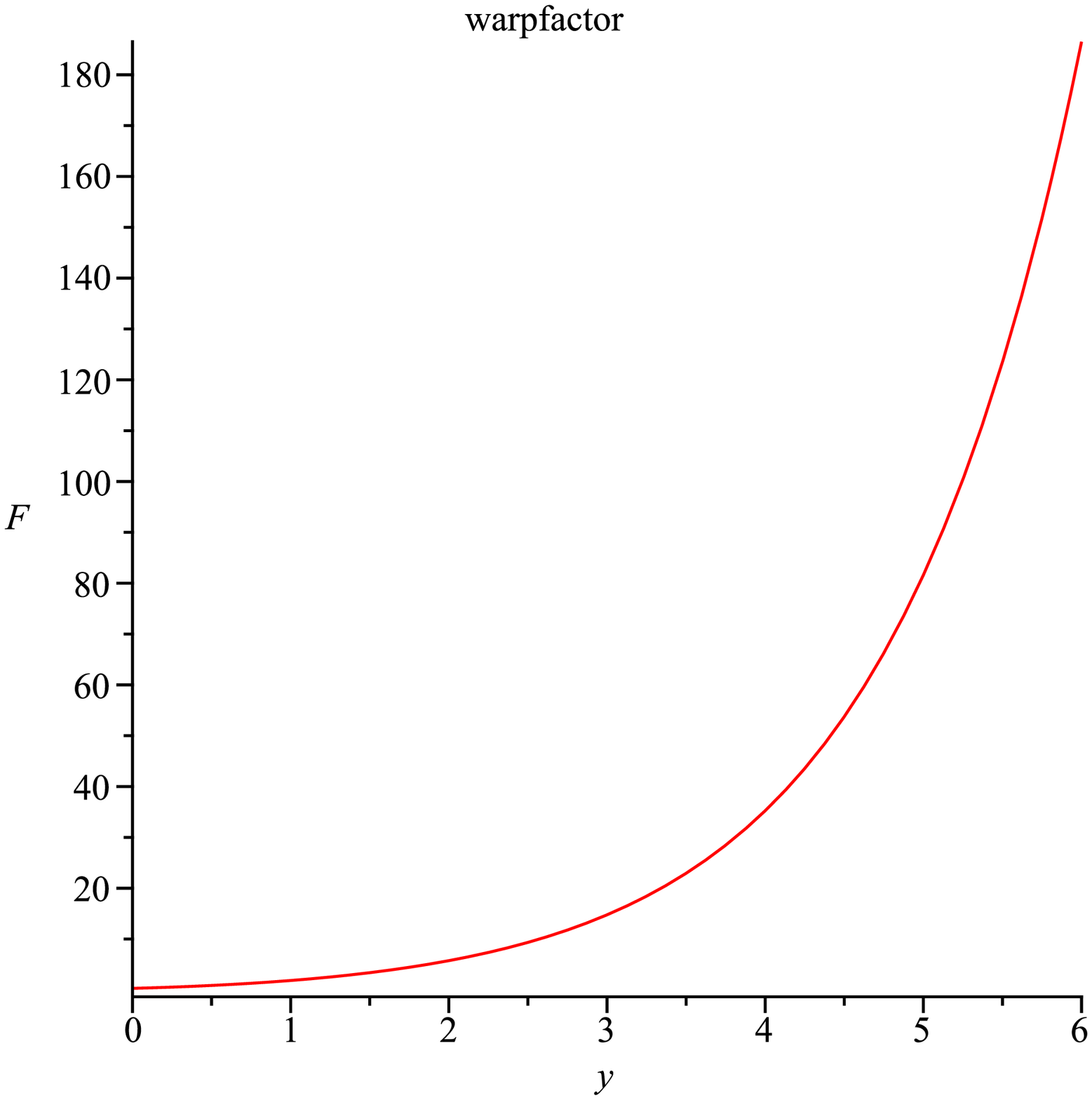}}
\vspace*{8pt}
\caption{Typical solutions of the warp factor F(y) in the case of an empty bulk. In the left solution, $\Lambda_5$ is taken  positive, where in the right solution $\Lambda_5$ is taken negative.\label{f2}}
\end{figure}

The evolution equations for $A(r,t)$ and $K(r,t)$ from the 5D Einstein equations are
\begin{equation}
\partial_{tt}A=e^A\Bigl(2\partial_{rr}A+\frac{3}{2}(\partial_rA)^2+3c_1\Bigr),\label{eqn11}
\end{equation}
\begin{equation}
\partial_{tt}K=\partial_tK\partial_tA+2e^{A}\Bigl(3c_1K+\partial_{rr}K+\frac{3}{4}K(\partial_rA)^2-\partial_rA\partial_rK\Bigr).\label{eqn12}
\end{equation}
These equations determine the evolution of the brane.
We will compare these equations with the induced 4D equations in the next chapter.
There is also a constraint equation, i.e.,
\begin{equation}
(\partial_r A)^2=\frac{4}{3}\frac{\partial_r K\partial_r A}{K}-\frac{2}{3}\frac{\partial_{rr}K}{K}-c_1.\label{eqn13}
\end{equation}
Constraint systems appear whenever a theory has gauge symmetry, in general relativity equivalent to coordinate change. Gauge invariance implies here general covariance under
coordinate transformations. In terms of fields on a space time  one says that the theory has redundancy and the constraint equations are preserved in time\cite{Bojo}.
\section{ The Field Equations on the Brane}
Following the review of R. Maartens\cite{Roy}, we have the induced field equations on the brane
\begin{equation}
{^{(4)}G}_{\mu\nu}=-\Lambda_{eff}{^{(4)}g}_{\mu\nu}+\kappa_4^2 {^{(4)}T}_{\mu\nu}+\kappa_5^4{\cal S}_{\mu\nu}-{\cal E}_{\mu\nu}
+\frac{2}{3}\kappa_5^4{\cal F}_{\mu\nu},\label{eqn14}
\end{equation}
where $\Lambda_{eff}=\frac{1}{2}(\Lambda_5+\kappa_4^2\lambda_4)=\frac{1}{2}(\Lambda_5+\frac{1}{6}\kappa_5^4\lambda_4^2)$,
 $\lambda_4$ is the vacuum energy in the brane (brane tension) and ${\cal F}_{\mu\nu}$ the energy-stress tensor contribution from the bulk scalar field.
The first correction term ${\cal S}_{\mu\nu}$ is  the quadratic term in the
energy-momentum tensor arising from the extrinsic curvature terms in the projected Einstein tensor
\begin{equation}
{\cal S}_{\mu\nu}=\frac{1}{12}{^{(4)}T}{^{(4)}T}_{\mu\nu}-\frac{1}{4}{^{(4)}T}_{\mu\alpha}{^{(4)}T}^\alpha_\nu
+\frac{1}{24}{^{(4)}g}_{\mu\nu}\Bigl[3{^{(4)}T}_{\alpha\beta}{^{(4)}T}^{\alpha\beta}-{^{(4)}T}^2\Bigr].\label{eqn15}
\end{equation}
The second correction term ${\cal E_{\mu\nu}}$ is given by
\begin{equation}
{\cal E}_{\mu\nu}={^{(5)}C}_{\alpha\gamma\beta\delta}n^\gamma n^\delta {^{(4)}g}_\mu^\alpha {^{(4)}g}_\nu^\beta \label{eqn16}
\end{equation}
and is a part of the 5D Weyl tensor and carries information of the gravitational field outside the brane and is constrained by the motion of the matter on the brane, i.e., the Codazzi equation
\begin{equation}
{^{(4)}\nabla}_\mu K^\mu_\nu-{^{(4)}\nabla}_\nu K={^{(5)}R}_{\mu\rho}{^{(4)}g}_\nu^\mu n^\rho,\label{eqn17}
\end{equation}
with $n^\mu$ the unit normal vector on the brane and $K_{\mu\nu}={^{(4)}g}_\mu^\rho {^{(5)}\nabla}_\rho n_\nu$
the extrinsic curvature. The induced metric on the brane is given by
${^{(4)}g}_{\mu\nu}={^{(5)}g}_{\mu\nu}-n_\mu n_\nu$.
${\cal E}_{\mu\nu}$ represents the 5D graviton effects, i.e., Kaluza-Klein modes in the linearized theory\cite{Ran1,Ran2}.
Integrating the 5D Einstein equations and using the Israel-Darmois junction conditions, one obtains for the extrinsic  curvature
\begin{equation}
K_{\mu\nu}=-\frac{1}{2}\kappa_5^2\Bigl({^{(4)}T}_{\mu\nu}+\frac{1}{3}(\lambda_4-{^{(4)}T}) {^{(4)}g}_{\mu\nu}\Bigr).\label{eqn18}
\end{equation}
One then obtains from Eq.(\ref{eqn3}),(17) and (18)
\begin{equation}
{^{(4)}\nabla^\nu} {^{(4)}T}_{\mu\nu}=-2 {^{(5)}T}_{\alpha\beta} n^\alpha g^\beta_\mu.\label{eqn19}
\end{equation}
It represents the exchange of energy-momentum between the bulk and the brane.
From the conservation equation ${^{(4)}\nabla^\nu}{^{(4)}G}_{\mu\nu}=0$,we then obtains
\begin{equation}
\kappa_4^2{^{(4)}\nabla^\nu}{^{(4)}T}_{\mu\nu}-{^{(4)}\nabla^\nu}{\cal E}_{\mu\nu}+\frac{6\kappa_4^2}{\lambda_4}{^{(4)}\nabla^\nu} {\cal S}_{\mu\nu}+4\frac{\kappa_4^2}{\lambda_4}
{^{(4)}\nabla^\nu}{\cal F}_{\mu\nu}=0.\label{eqn20}
\end{equation}
If we take an empty bulk, we have simply ${^{(4)}\nabla^\nu}{^{(4)}T}_{\mu\nu}=0$ and so
\begin{equation}
{^{(4)}\nabla^\nu} {{\cal E}}_{\mu\nu}=\kappa_5^4{^{(4)}\nabla^\nu} {{\cal S}}_{\mu\nu}.\label{eqn21}
\end{equation}
In the static case of the space time Eq.(\ref{eqn4}), one can evaluate the equations for $K$ and $A$ and the gauge fields $X$ and $P$. In order to compare the change in behavior of these equations
with respect to 4D counterpart equations, we take the same combinations of the components of  the Einstein equations as in the 4D case. The result is
\begin{eqnarray}
\partial_{rr}K-\frac{3}{7}\partial_rK\partial_r A+\frac{9}{28}K(\partial_rA)^2=\cr
\frac{3}{7}\kappa_4^2\Bigl[-\frac{3}{4}\beta K(X^2-\eta^2)^2+\frac{e^{2A}}{Ke^2}(\partial_r P)^2-2\frac{X^2P^2}{K}e^{2A}\Bigr] -2\frac{4}{7}K\Lambda_{eff}\cr
+\frac{1}{896}\kappa_5^4\Bigl[\frac{8\beta(X^2-\eta^2)^2}{e^2K}(\partial_r P)^2e^{2A}-\frac{32X^2P^2}{K}(\partial_rX)^2e^{2A}+48K(\partial_rX)^4\cr
+\frac{256X^2P^2}{e^2K^3}(\partial_rP)^2e^{4A}+\frac{64}{e^2K}(\partial_rX)^2(\partial_rP)^2e^{2A}+\frac{144}{e^4K^3}(\partial_rP)^4e^{4A}\cr
-3\beta^2K(X^2-\eta^2)^4+\frac{112X^4P^4}{K^3}e^{4A}-\frac{16\beta X^2P^2(X^2-\eta^2)^2}{K}e^{2A}\Bigr],\label{eqn22}
\end{eqnarray}
\begin{eqnarray}
\partial_{rr}A+\frac{1}{2}\frac{\partial_r K\partial_rA}{K}+\frac{3}{8}(\partial_r A)^2+\frac{1}{2}\frac{\partial_{rr}K}{K}=\cr
\frac{3}{2}\kappa_4^2\Bigl[\frac{(\partial_r P)^2}{e^2 K^2}e^{2A}-\frac{1}{4}\beta(X^2-\eta^2)^2\Bigr]-3\Lambda_{eff}\cr
+\frac{1}{256}\kappa_5^4\Bigl[\frac{8\beta(X^2-\eta^2)^2}{e^2K^2}e^{2A}(\partial_r P)^2+\frac{32X^2P^2}{K^2}e^{2A}(\partial_rX)^2+16(\partial_r X)^4 \cr
+\frac{64X^2P^2}{e^2K^4}e^{4A}(\partial_r P)^2+\frac{64}{e^2K^2}(\partial_rX)^2(\partial_rP)^2 e^{2A}\cr
+\frac{48}{e^4K^4}(\partial_rP)^4e^{4A}-\beta^2(X^2-\eta^2)^4+\frac{16X^4P^4}{K^4}e^{4A}\Bigr],\label{eqn23}
\end{eqnarray}
\begin{eqnarray}
\partial_{rr} X=-\frac{\partial_r X\partial_r K}{K}+\frac{e^{2A}XP^2}{K^2}+\frac{1}{2}\beta X(X^2-\eta^2), \cr
\partial_{rr}P=\frac{\partial_r P\partial_r K}{K}-2P_rA_r+e^2X^2P.\label{eqn24}
\end{eqnarray}
We see that the left hand sides of the Einstein equations are different with respect to the ones in the 4D case. This is a consequence  of  the ${\cal E} _{\mu\nu}$
term entering the equations. On the right hand sides we recognize the $\kappa_4^2$-terms of  the 4D case [ see Eq.(\ref{app7})and (\ref{app8})].
When we substitute  the constraint equation of $(A_r)^2$ (Eq.\ref{eqn13}) of
the 5D equations (valid for all t) and using the same notations as is the 4D case ( see Appendix), i.e.,
\begin{eqnarray}
\Theta_1\equiv K\partial_r A, \qquad  \Theta_2\equiv \partial_r K,\cr
 T_{\mu\nu}=\sigma \hat{k}_t\hat{k}_t+\varrho_z\hat{k}_z\hat{k}_z+\varrho_\varphi \hat{k}_\varphi\hat{k}_\varphi+\varrho_r\hat{k}_r\hat{k}_r,\cr
 {\cal S}_{\mu\nu}=\xi_t \hat{k}_t\hat{k}_t+\xi_z\hat{k}_z\hat{k}_z+\xi_\varphi \hat{k}_\varphi\hat{k}_\varphi+\xi_r\hat{k}_r\hat{k}_r,\label{eqn25}
\end{eqnarray}

then the equations for the metric components become
\begin{eqnarray}
\partial_r\Theta_2=\frac{6}{11} K\Bigl[\kappa_4^2(3\varrho_r-2\sigma +\varrho_\varphi )+\kappa_5^4(3\xi_r-2\xi_t +\xi_\varphi )+\frac{3c_1}{4}-6\Lambda_{eff}\Bigr].\label{eqn26}
\end{eqnarray}
\begin{eqnarray}
4\partial_r\Theta_1+\partial_r\Theta_2=
6K\Bigl[\kappa_4^2(\varrho_r +\varrho_\varphi) +\kappa_5^4(\xi_r +\xi_\varphi)+\frac{c_1}{4}-2\Lambda_{eff}\Bigr].\label{eqn27}
\end{eqnarray}

In the special case of $c_1=8\Lambda_{eff}$, we see on the right hand side the same combinations of the energy-momentum tensor components.
We will use these equations to evaluate the angle deficit in our model.

If we would take $ \kappa_4^2 = \frac{1}{6}\lambda_4 \kappa_5^4 $, we then  have 5 parameters for the model, i.e., $\beta , e, \eta, \lambda_4$
and $\kappa_5$.
The effective field equations are supplemented by the 5D equations and the conservation of stress-energy.

We can compare the numerical solution of the 4D-brane equations with the "classical" Nielsen-Olesen solution of the appendix A. In figure 3 we plotted the solution with the same initial
conditions and parameters. We used a standard routine using solely initial conditions. In order to find an "acceptable" solution, one has to fine tune the initial values.\cite{Dyer}.
We observe that only $K$ behaves differently.

\begin{figure}
\centerline{
\includegraphics[width=4cm]{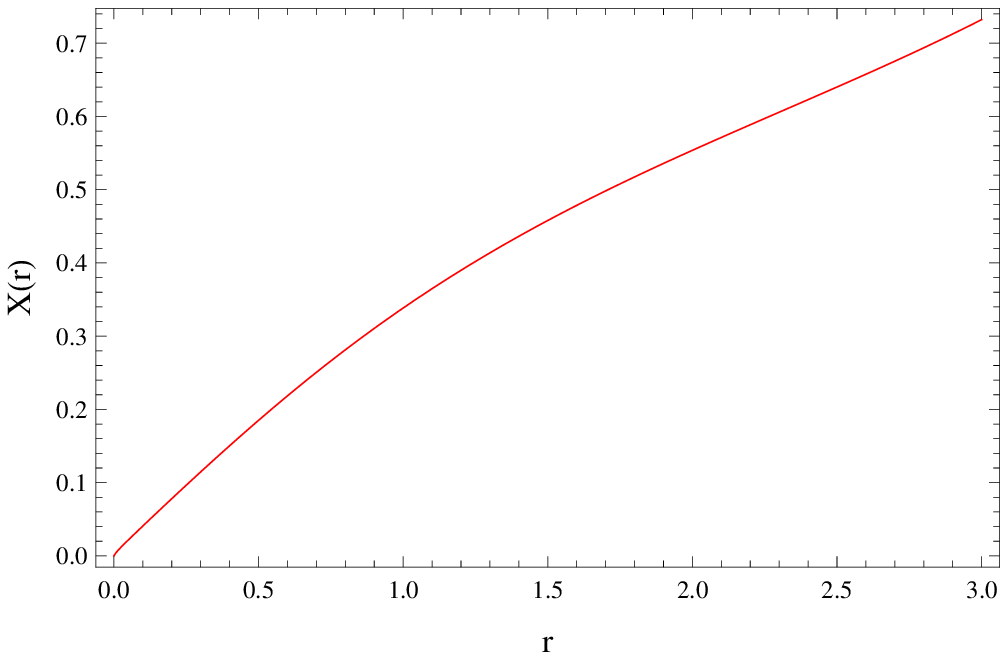}
\includegraphics[width=4cm]{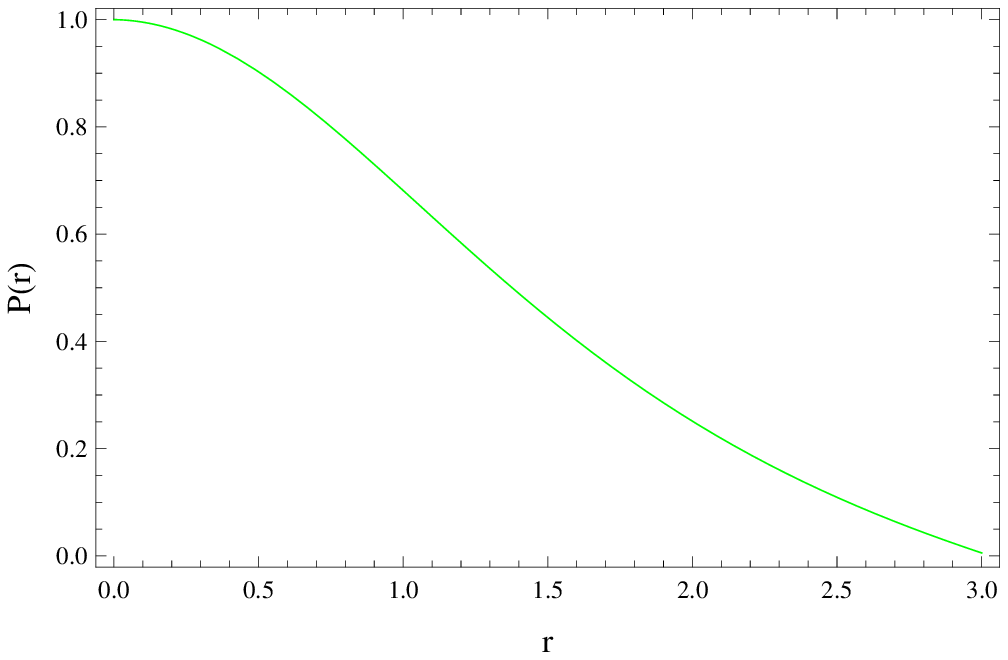}
\includegraphics[width=4cm]{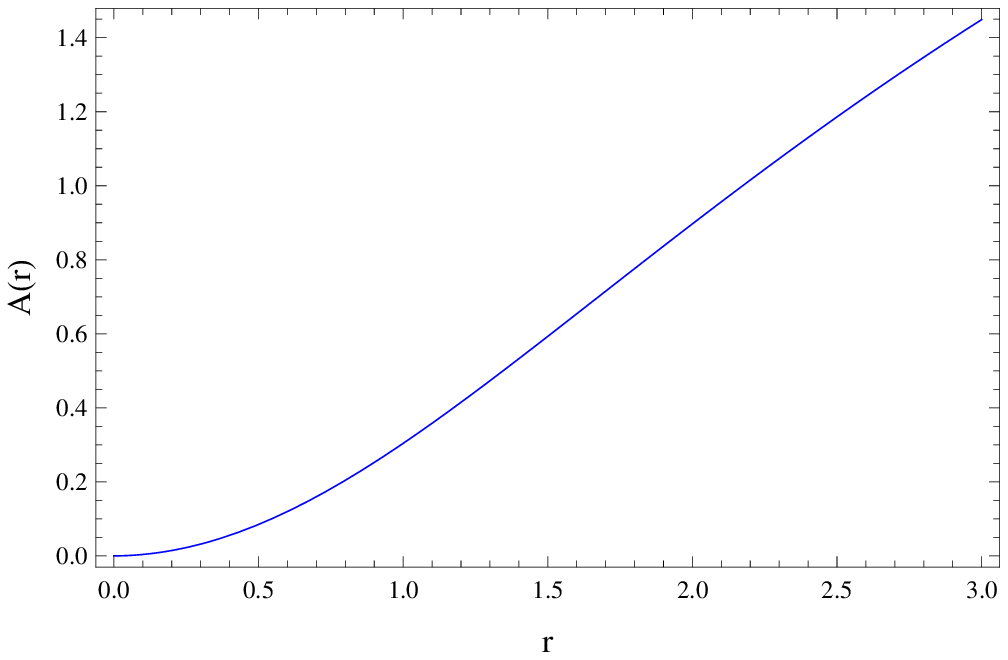}
\includegraphics[width=4cm]{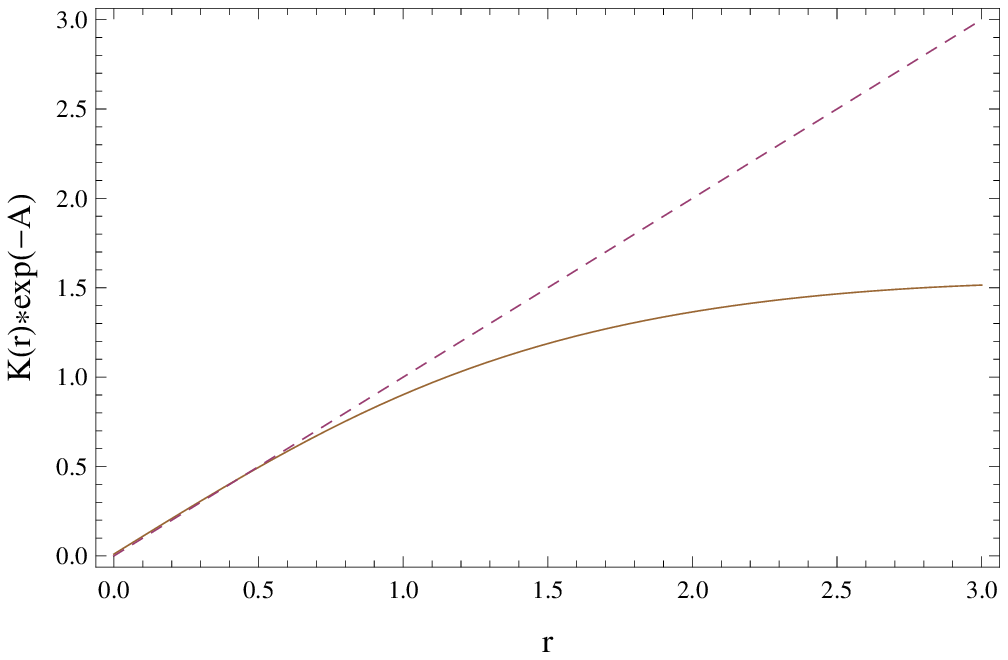}}
\centerline{
\includegraphics[width=4cm]{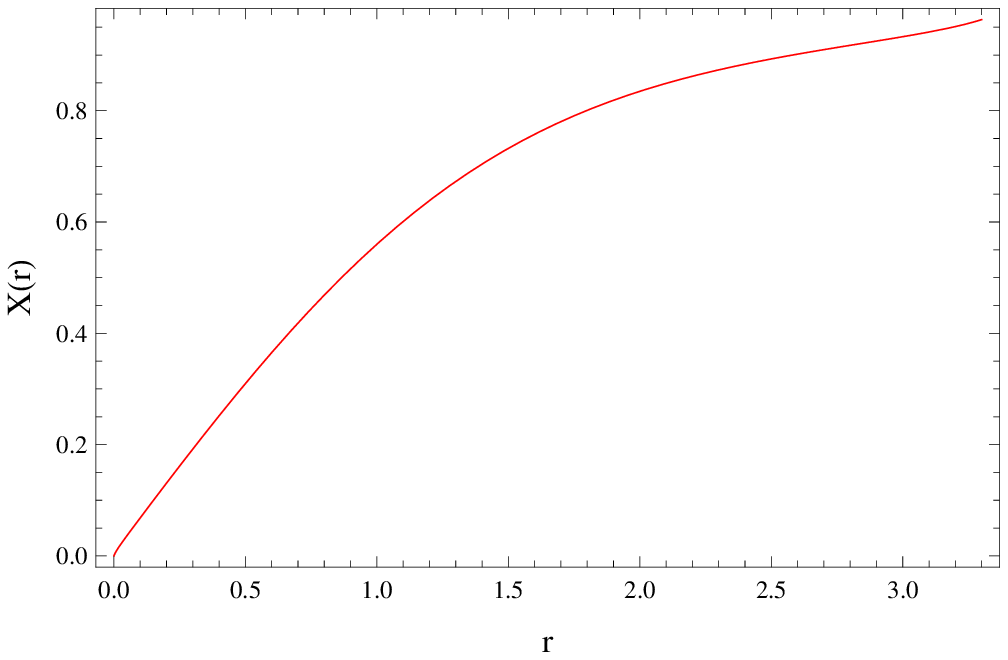}
\includegraphics[width=4cm]{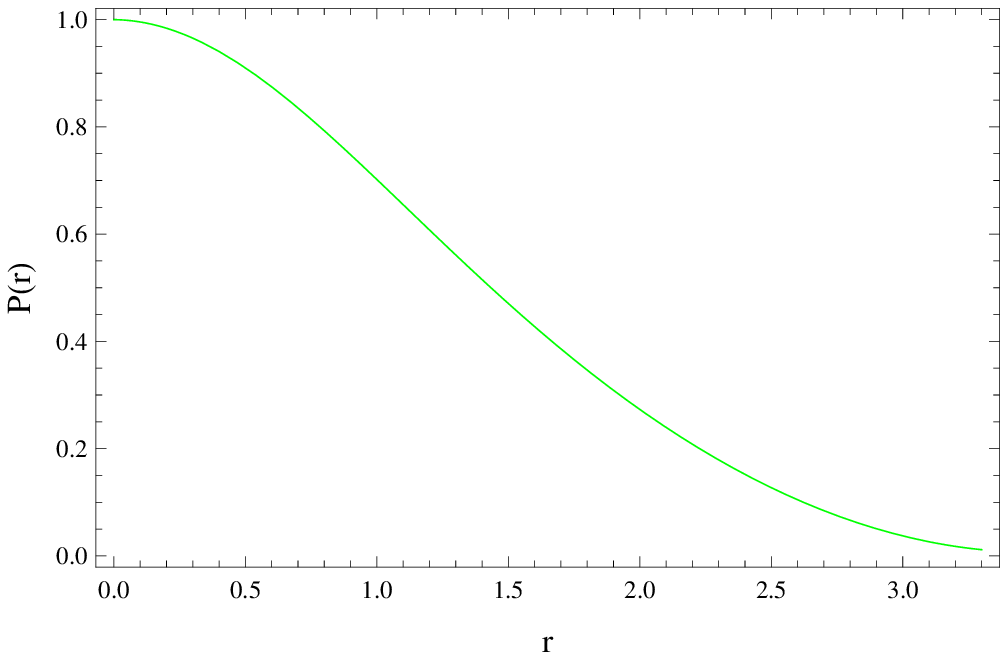}
\includegraphics[width=4cm]{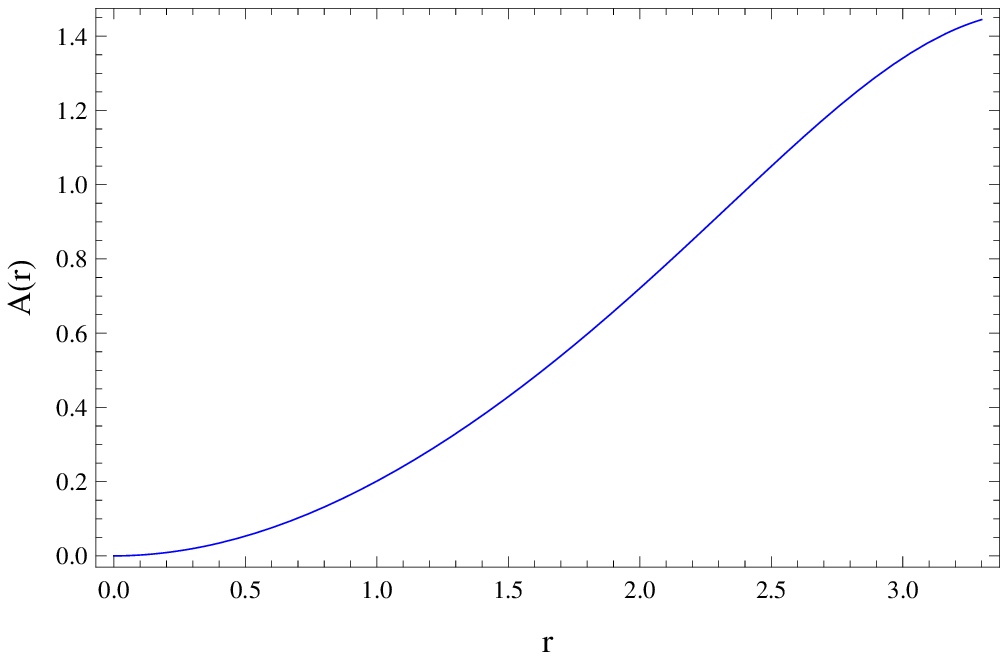}
\includegraphics[width=4cm]{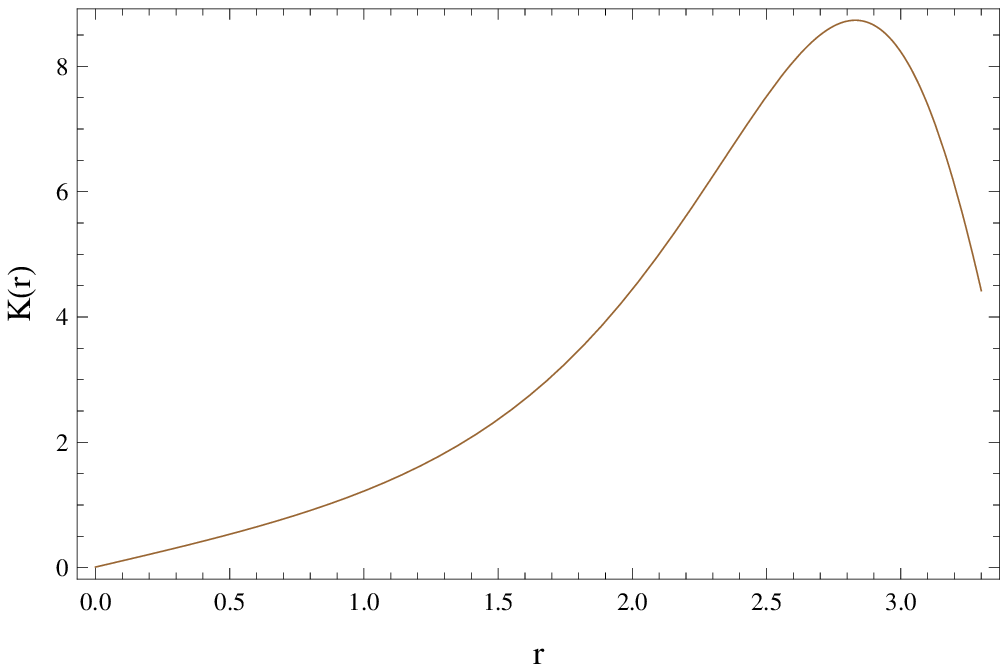}}
\vspace*{8pt}
\caption{Two characteristic solution of the brane induces U(1) gauge string. We used a standard
"shooting" routine.  We observe the same behavior of $\Phi$ and $P$, but a significant different behavior of $K$ and $K.e^{-A}$. The dashed line represents the Minkowski space time.\label{f3}}
\end{figure}

We can also use a two point boundary condition routine. In figures 4 and 5 we plotted the results.
\begin{figure}
\centerline{
\includegraphics[width=4cm]{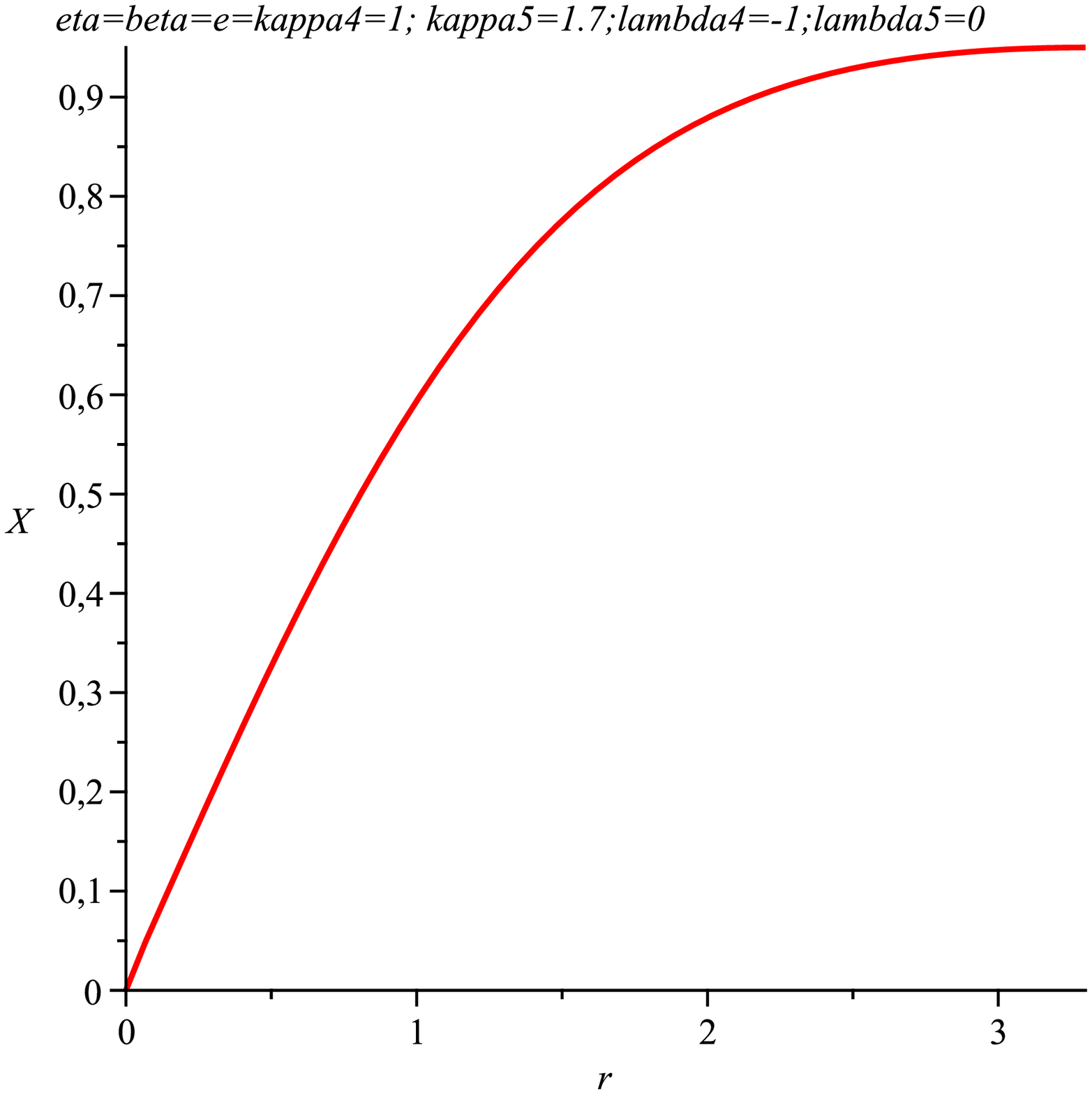}
\includegraphics[width=4cm]{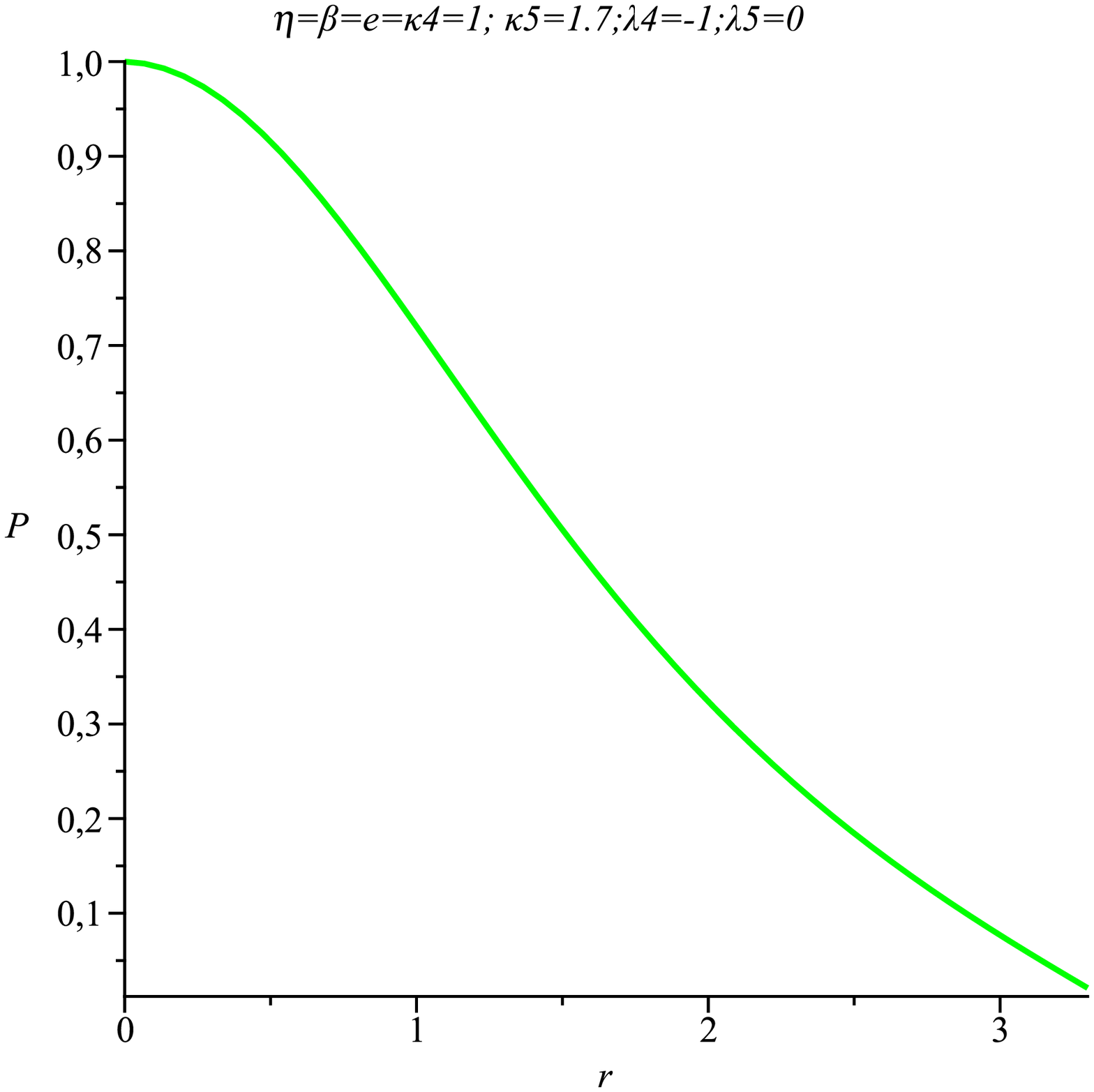}
\includegraphics[width=4cm]{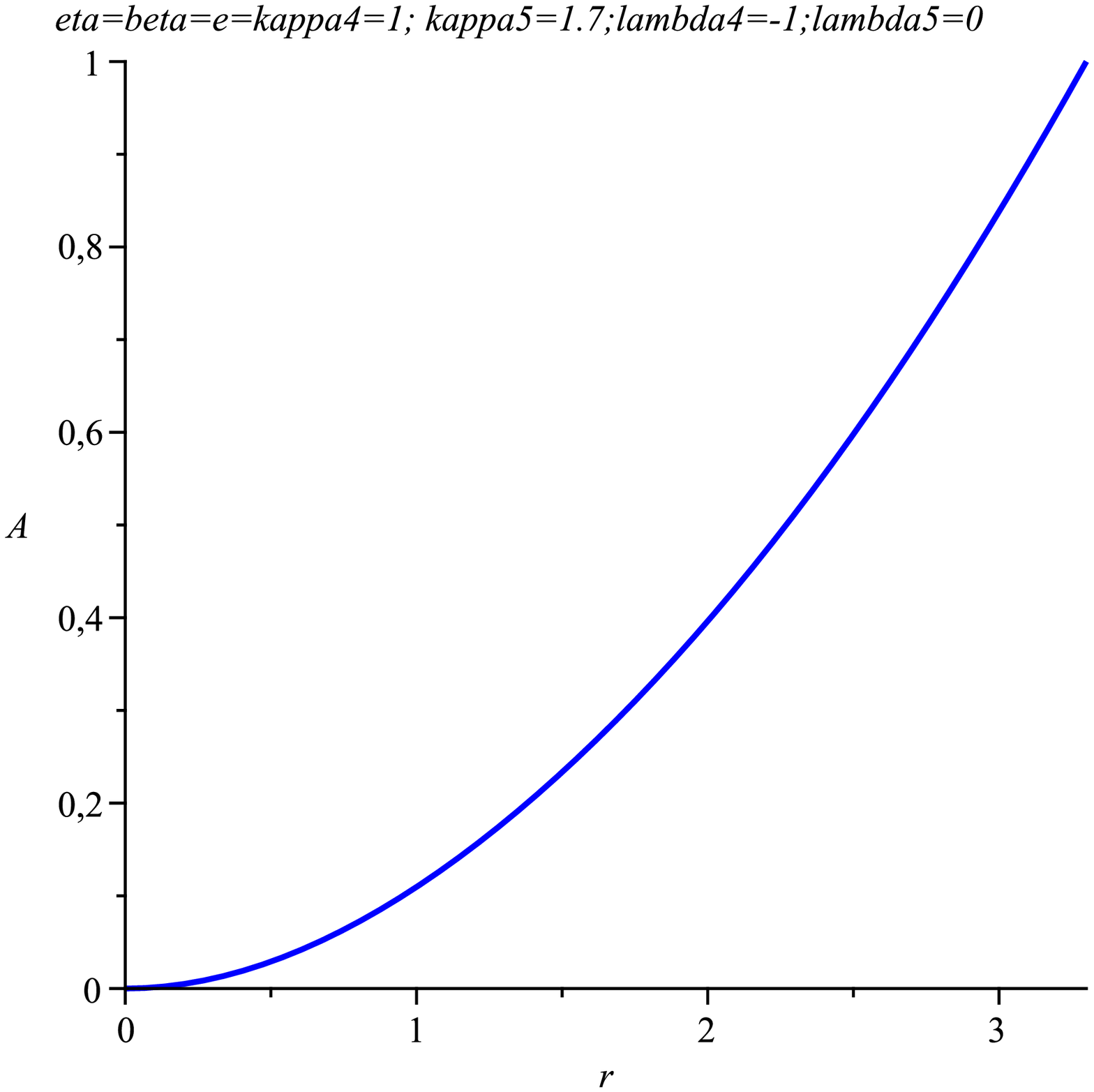}
\includegraphics[width=4cm]{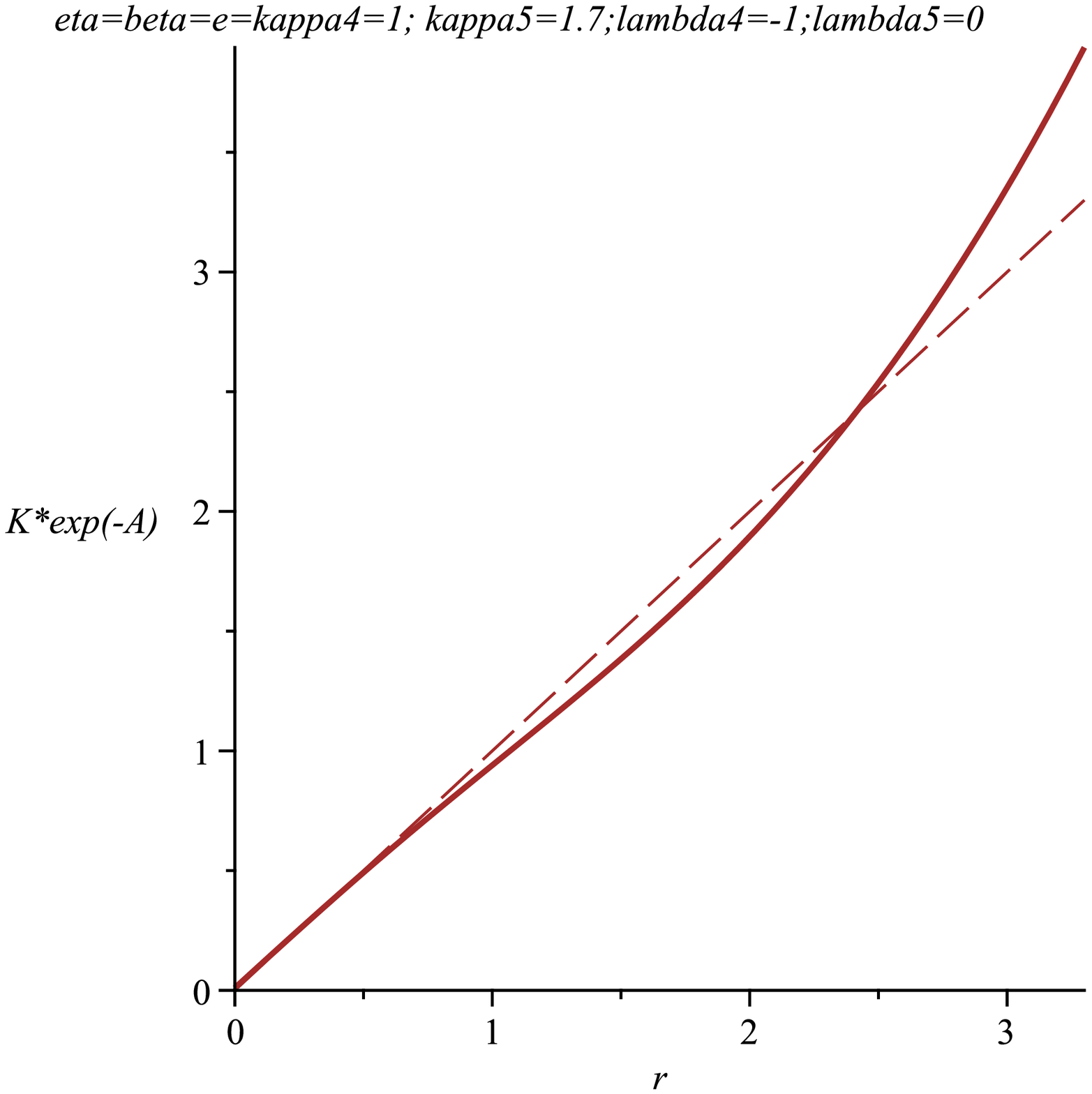}}
\vspace*{8pt}
\caption{As figure 3. Typical numerical solution of the brane induces U(1) gauge string using a two point boundary value routine.  The behavior of $e^{-A}K$  is non-regular.\label{f4}}
\end{figure}

\begin{figure}
\centerline{
\includegraphics[width=5cm]{G16.eps}
\includegraphics[width=5cm]{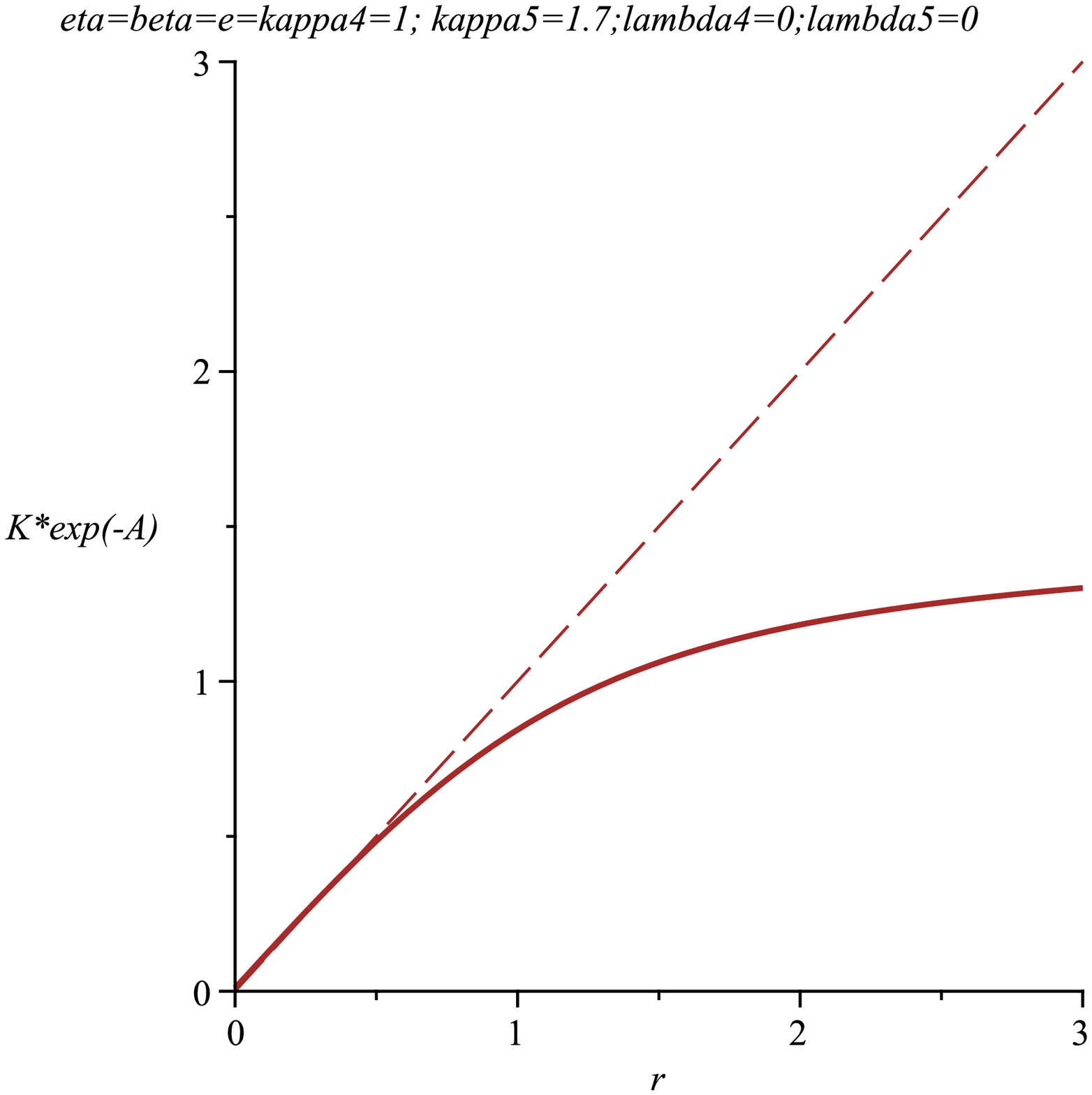}
\includegraphics[width=5cm]{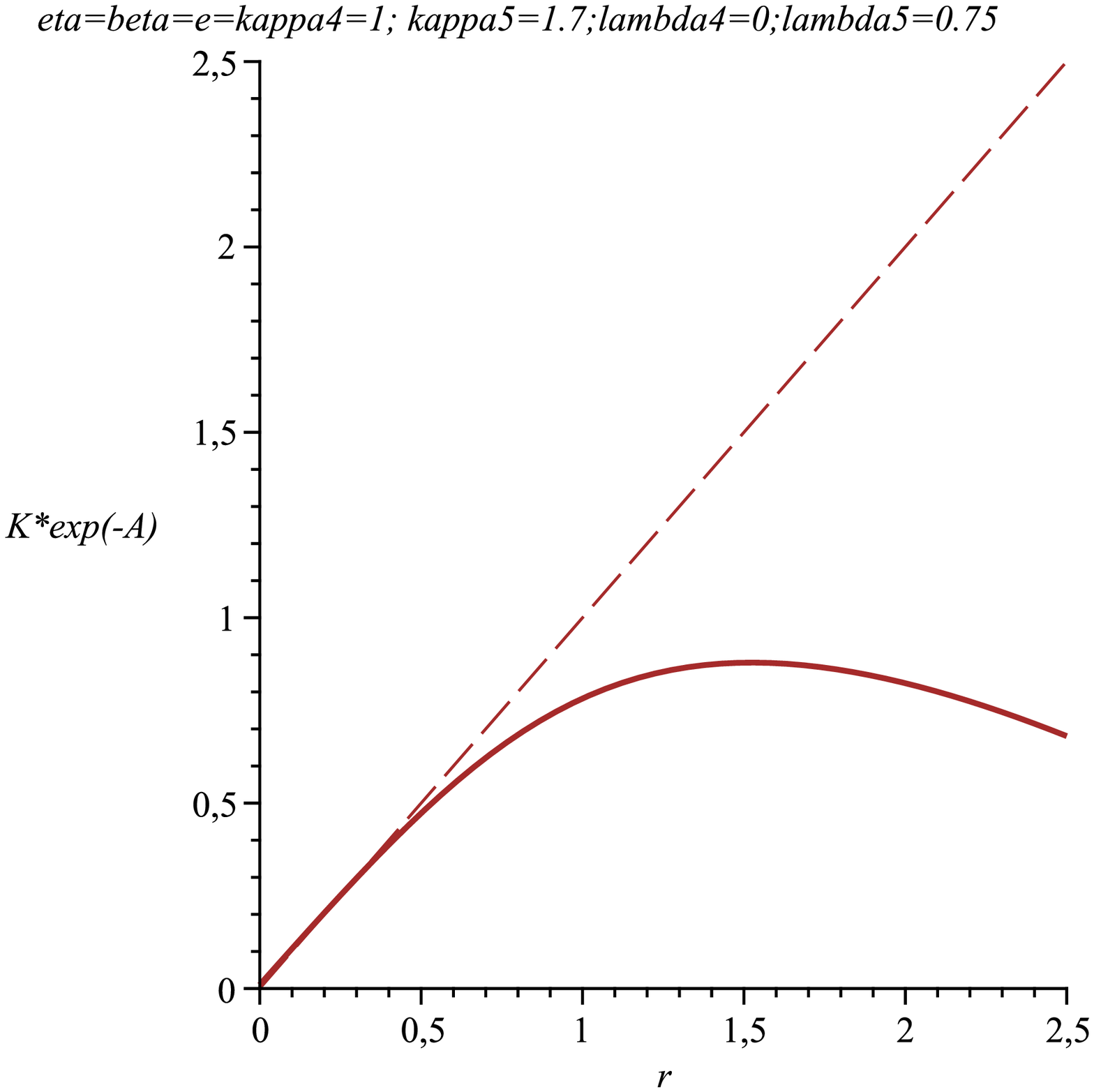}}
\centerline{
\includegraphics[width=5cm]{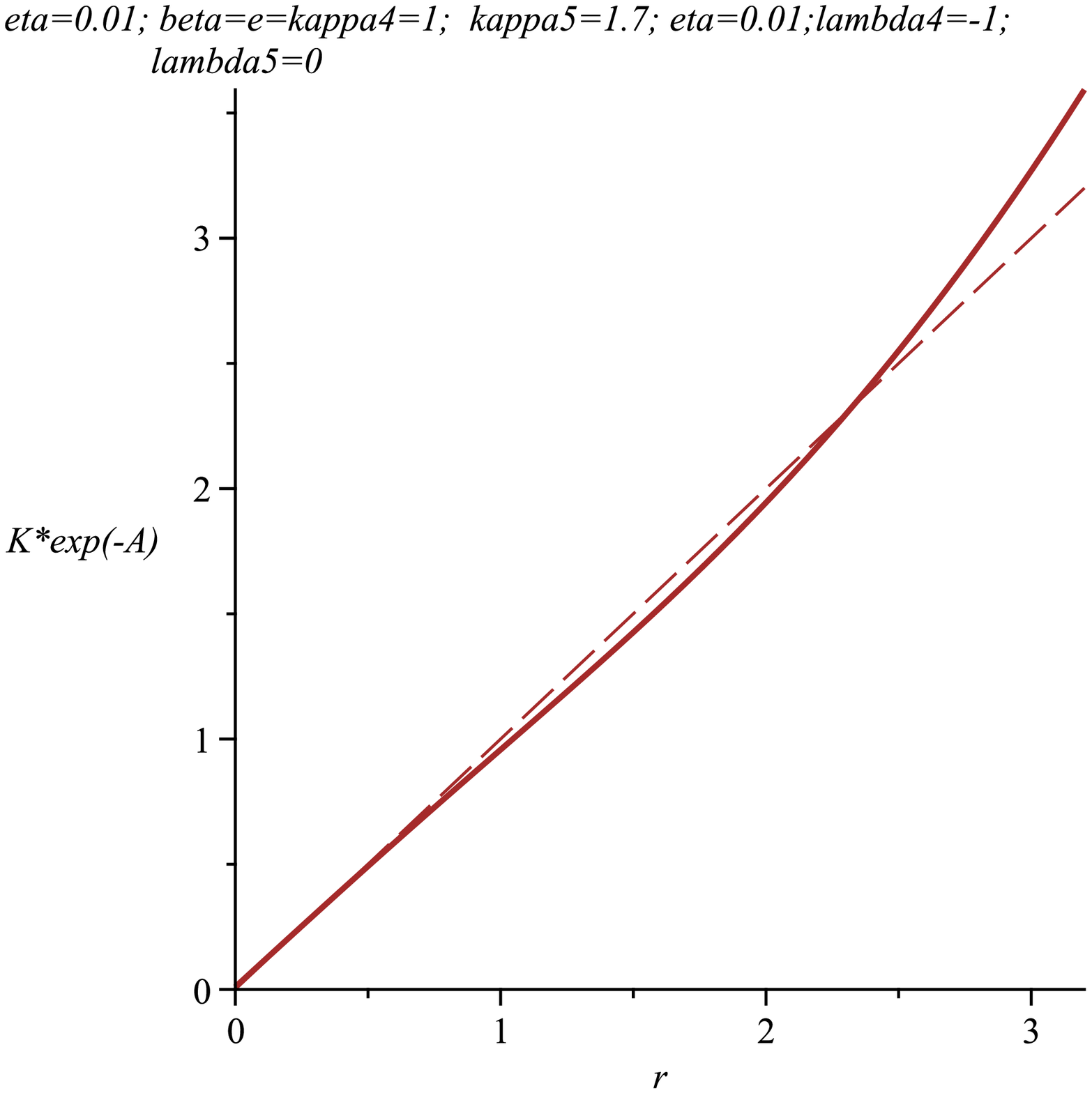}
\includegraphics[width=5cm]{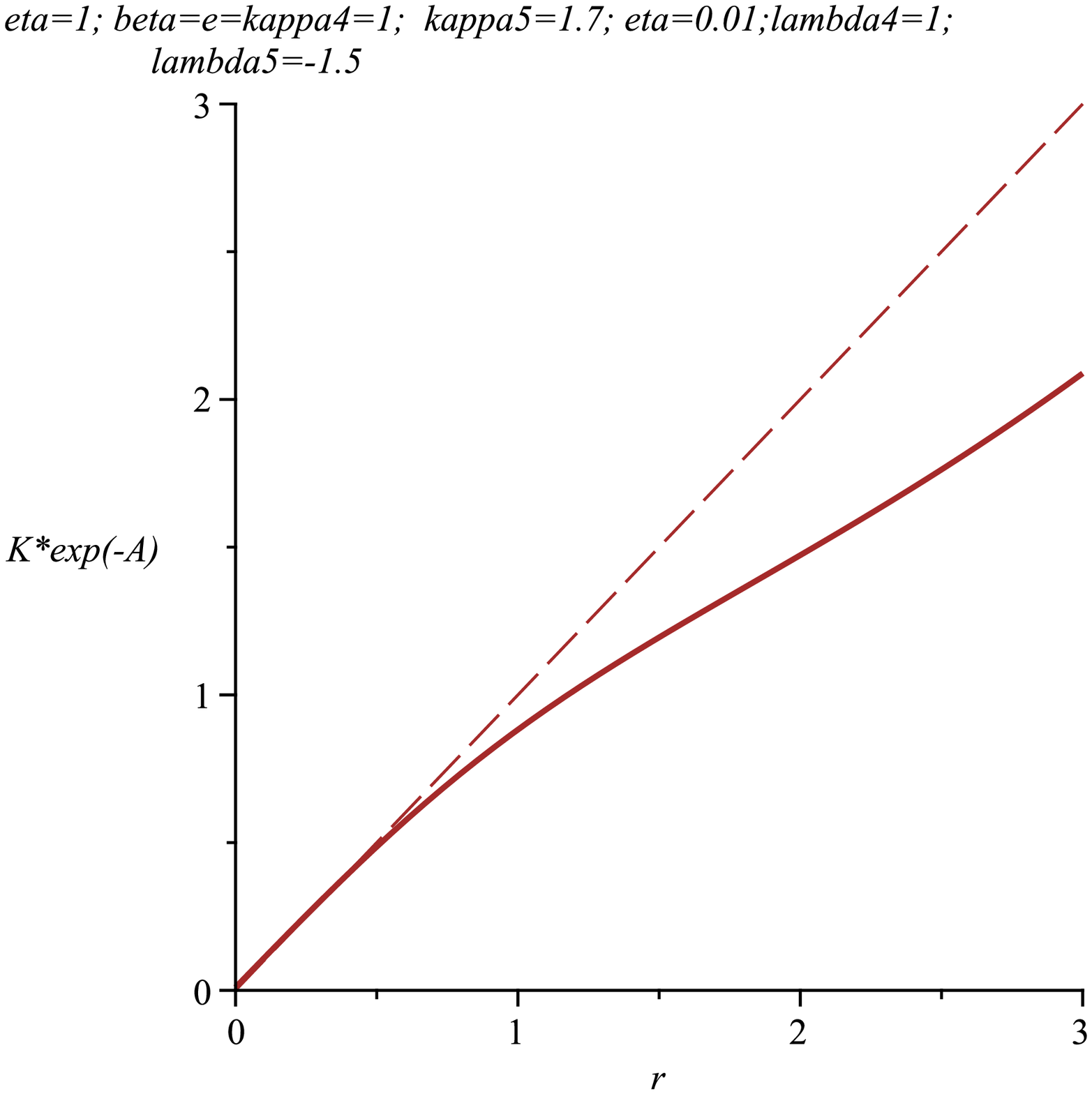}}
\vspace*{8pt}
\caption{Numerical solution of the metric component $e^{-A}K$ for different values of the brane and bulk cosmological constants.
The last one represents the most realistic situation, where $\lambda_4 >0$ and $\Lambda_5 <0$. The behavior don't change significantly with increasing $\eta$.\label{f5}}
\end{figure}

This system should be numerically investigated with a more advanced integration model, specially in the full time dependent situation.
This is currently under study by the authors.

\section{ Analysis of the Angle Deficit}
The angle deficit can be calculated for a class of static translational symmetric space times which are asymptotically Minkowski minus a wedge.
If we denote with $l$ the length of an orbit of $\Bigl(\frac{\partial}{\partial\varphi}\Bigr)^a$ in the brane, then the angle deficit is given
by\cite{Garf,Vil,Ford Vil}
\begin{equation}
(2\pi -\Delta\theta )=\lim_{r\rightarrow\infty}\frac{d l}{dr},\label{eqn28}
\end{equation}
with
\begin{equation}
l=\int_0^{2\pi}\sqrt{g_{ab}\Bigl(\frac{\partial}{\partial\varphi}\Bigr)^a\Bigl(\frac{\partial}{\partial\varphi}\Bigr)^b} d\varphi =2\pi e^{-A}K.\label{eqn29}
\end{equation}
Using boundary conditions at the axis, we obtain
\begin{eqnarray}
\Delta\theta =-2\pi\int_{0}^{\infty}\frac{d^2}{dr^2}(e^{-A}K)dr \cr =-2\pi\int_{0}^{\infty} \Bigl[e^{-A}\Bigl(\frac{d\Theta_2}{dr}-\frac{d\Theta_1}{dr}\Bigr)-\frac{\Theta_1}{K}e^{-A}(\Theta_2-\Theta_1)\Bigr]dr.\label{eqn30}
\end{eqnarray}

Let us now evaluate $\frac{d}{dr}(\kappa_4^2K^2\varrho_r)$, using the conservation of stress energy ( see appendix Eq.(\ref{app15})) and the two field equations Eq.(\ref{eqn26})-(27) for A and K.

\begin{eqnarray}
\frac{d}{dr}(\kappa_4^2K^2\varrho_r)=\kappa_4^2(2K\frac{dK}{dr}\varrho_r+K^2\frac{d\varrho_r}{dr})=\kappa_4^2K\Bigl(\varrho_r\frac{dK}{dr}+\frac{d}{dr}(K\varrho_r)\Bigr)\cr
=\kappa_4^2K\Bigl(\Theta_2\varrho_r+\Theta_2\varrho_\varphi -\Theta_1(\sigma+\varrho_\varphi )\Bigr)=\kappa_4^2K\Bigl(\Theta_2(\varrho_r+\varrho_\varphi)-\Theta_1(\sigma +\varrho_\varphi)\Bigr)\cr
=\Theta_2(\frac{2}{3}\frac{d\Theta_1}{dr}+\frac{1}{6}\frac{d\Theta_2}{dr})+\frac{1}{2}\kappa_4^2K(-2\sigma -2\varrho_\varphi )\cr
=\Theta_2(\frac{2}{3}\frac{d\Theta_1}{dr}+\frac{1}{6}\frac{d\Theta_2}{dr})+\kappa_4^2K\Theta_1\Bigl(\frac{1}{2}(-2\sigma+3\varrho_r
+\varrho_\varphi )+(-\frac{3}{2}\varrho_r-\frac{3}{2}\varrho_\varphi )\Bigr)\cr
=\Theta_2(\frac{2}{3}\frac{d\Theta_1}{dr}+\frac{1}{6}\frac{d\Theta_2}{dr})+\frac{11}{12}\Theta_1\frac{d\Theta_2}{dr}-\frac{3}{2}\Theta_1 (\frac{2}{3}\frac{d\Theta_1}{dr}+\frac{1}{6}\frac{d\Theta_2}{dr})\cr
=\Theta_2(\frac{2}{3}\frac{d\Theta_1}{dr}+\frac{1}{6}\frac{d\Theta_2}{dr})+\frac{11}{12}\Theta_1\frac{d\Theta_2}{dr}-\Theta_1\frac{d\Theta_1}{dr}-\frac{1}{4}\Theta_1\frac{d\Theta_2}{dr}\cr
=\frac{2}{3}\Theta_2\frac{d\Theta_1}{dr}+\frac{1}{6}\Theta_2\frac{d\Theta_2}{dr}-\Theta_1\frac{d\Theta_1}{dr}+\frac{2}{3}\Theta_1\frac{d\Theta_2}{dr}\cr
=\frac{d}{dr}\Bigl[\frac{2}{3}\Theta_1\Theta_2+\frac{1}{12}\Theta_2^2-\frac{1}{2}\Theta_1^2\Bigr].\label{eqn31}
\end{eqnarray}
Boundary conditions at the axis then imply
\begin{equation}
\frac{2}{3}\Theta_1\Theta_2+\frac{1}{12}\Theta_2^2-\frac{1}{2}\Theta_1^2=\kappa_4^2K^2\varrho_r.\label{eqn32}
\end{equation}

We used the case where  $c1=8\Lambda_{eff}$. Further, we used that ${\cal S}_{\mu\nu} \sim(T_{\mu\nu})^2$, so
\begin{equation}
\frac{|\kappa_4^2{\cal S}_{\mu\nu}/ \lambda_4|}{|\kappa_4^2 {^{(4)}T}_{\mu\nu}|}\sim\frac{|^{(4)}T_{\mu\nu}|}{\lambda_4}.\label{eqn33}
\end{equation}
Further we assumed $|{^{(4)}T}_{\mu\nu}| <<\lambda_4$\cite{Roy}.

As in the 4D case we assume that $$\lim_{r\rightarrow\infty} K^2\sigma \rightarrow 0$$ and that $\sigma >|\varrho_r|>|\varrho_\varphi| $.
From the field equations we then have that $\Theta_1$ and $\Theta_2$ approach constant values $k_1,k_2$ as $r \rightarrow \infty$.   We then obtain from Eq.(\ref{eqn32})
\begin{equation}
\bar \Theta_1=\frac{1}{-4\pm \sqrt{22}}\bar \Theta_2,\label{eqn34}
\end{equation}
where we denote with $\bar \Theta_1$ and $\bar \Theta_2$ the asymptotic values.
So we have for  $\partial_r\bar A$
\begin{equation}
\partial_r\bar A =\frac{\partial_r \bar K}{\bar K(-4\pm\sqrt{22})}.\label{eqn35}
\end{equation}
The solutions for  $\bar K$ and $\bar A$ are then
\begin{eqnarray}
\bar K=k_2 r + a_2, \qquad \bar A =\frac{\ln(k_2 r+a_2)}{-4\pm\sqrt{22}}+a_0,\label{eqn36}
\end{eqnarray}
and the space time becomes
\begin{equation}
ds^2=e^{a_0}(k_2 r+a_2)^{\frac{1}{-4\pm\sqrt{22}}}[-dt^2+dz^2]+dr^2+e^{-2a_0}(k_2 r+a_2)^{2+\frac{2}{4\mp\sqrt{22}}}d\varphi^2.\label{eqn37}
\end{equation}
Let us compare our relation Eq.(\ref{eqn32}) with the 4D counterpart of Eq.(\ref{app16}). The 4D solution $k_1=\bar K\partial_r \bar A=0$ is no option here ( see Eq.(\ref{app17})).
We have now two possibilities for the asymptotic space time: both non- Kasner-like and non-conical.

Combined with the warp factor $F(y)$, the  space time Eq.(\ref{eqn4})  becomes

\begin{equation}
ds^2=F(y)\Bigl[e^{a_0}(k_2 r+a_2)^{1.448}[-dt^2+dz^2]+dr^2+e^{-2a_0}(k_2 r+a_2)^{-0,897}d\varphi^2\Bigr]+dy^2,\label{eqn38}
\end{equation}
or
\begin{equation}
ds^2=F(y)\Bigl[e^{a_0}(k_2 r+a_2)^{-0.115}[-dt^2+dz^2]+dr^2+e^{-2a_0}(k_2 r+a_2)^{2.23}d\varphi^2\Bigr]+dy^2,\label{eqn39}
\end{equation}
with $F(y)$ in the empty bulk situation given by Eq.(\ref{eqn10})
These solutions are in general un-physical. The behavior depends on the sign of $\frac{k_2}{a_2}$. Under less restrictive conditions, for example, $c_1\neq8\Lambda_{eff}$
and with special choices of the parameters, the numerical solutions show some regular behavior.

Now we can evaluate the angle deficit Eq.(\ref{eqn30}). We can make a linear combination of Eq.(\ref{eqn26}),(27) and (32) in order to isolate the term $e^{-A}K\sigma $ ( as in the 4D case):

\begin{eqnarray}
\Delta\theta=-2\pi\int_{0}^{\infty}e^{-A}\Bigl[\Bigl(\frac{5}{6}\frac{d\Theta_2}{dr}-\frac{1}{3}\frac{d\Theta_1}{dr}\Bigr) +\frac{1}{6}\frac{d\Theta_2}{dr}-\frac{2}{3}
\frac{d\Theta_1}{dr} - \frac{\Theta_1}{K}(\Theta_2-\Theta_1)\Bigr]dr \cr
=-2\pi\int_{0}^{\infty}e^{-A}\Bigl[\kappa_4^2 K(\varrho_r-\sigma )+\frac{1}{6}\frac{d\Theta_2}{dr}-\frac{2}{3}
\frac{d\Theta_1}{dr} - \frac{\Theta_1}{K}(\Theta_2-\Theta_1)\Bigr]dr \cr
=2\pi\kappa_4^2\int_{0}^{\infty}e^{-A}K\sigma dr-2\pi\int_{0}^{\infty}e^{-A}\Bigl[\frac{1}{K}\Bigl(\frac{2}{3}\Theta_1\Theta_2+\frac{1}{12}\Theta_2^2-\frac{1}{2}\Theta_1^2\Bigr)\cr
+\frac{1}{6}\frac{d\Theta_2}{dr}-\frac{2}{3}
\frac{d\Theta_1}{dr} - \frac{\Theta_1}{K}(\Theta_2-\Theta_1)\Bigr]dr \cr
=2\pi\kappa_4^2\int_{0}^{\infty}e^{-A}K\sigma dr-2\pi\int_{0}^{\infty}e^{-A}\Bigl[\frac{1}{K}\Bigl(-\frac{1}{3}\Theta_1\Theta_2+\frac{1}{12}\Theta_2^2+\frac{1}{2}\Theta_1^2\Bigr)\cr
+\frac{1}{6}\frac{d\Theta_2}{dr}-\frac{2}{3}\frac{d\Theta_1}{dr} \Bigr]dr\cr
=2\pi\kappa_4^2\int_{0}^{\infty}e^{-A}K\sigma dr-2\pi\int_{0}^{\infty}\Bigl[\frac{e^{-A}}{K}\Bigl(-\frac{1}{3}\Theta_1\Theta_2+\frac{1}{12}\Theta_2^2+\frac{1}{2}\Theta_1^2\Bigr)\cr
+\frac{1}{6}\Bigl(\frac{d}{dr}(e^{-A}\Theta_2)+\frac{e^{-A}\Theta_1\Theta_2}{K}\Bigr)-\frac{2}{3}\Bigl(\frac{d}{dr}(e^{-A}\Theta_1)+\frac{e^{-A}\Theta_1^2}{K}\bigr)\Bigr]dr\cr
=\kappa_4^2\mu-2\pi\int_{0}^{\infty}\Bigl[\frac{e^{-A}}{K}\Bigl(\frac{2}{3}\Theta_1\Theta_2-\frac{1}{6}\Theta_1^2+\frac{1}{12}\Theta_2^2\Bigr)\cr
+\frac{1}{6}\frac{d(e^{-A}\Theta_2)}{dr}-\frac{2}{3}\frac{d(e^{-A}\Theta_1)}{dr}\Bigr]dr.\label{eqn40}
\end{eqnarray}

The first term is again the linear energy density of the string ( see Eq.(\ref{app23})) and  is of order $\eta^2$.
The correction terms are in contrast with the 4D case, unbounded and will give chaotic results, as is seen in the numerical solutions of figure (5).
Only for positive brane tension and negative bulk tension (the 5D brane-world  preferred values) there seems to be a stable solution for $e^{-A}K$. However, this is
not a "classical" cosmic string situation.

\section{Conclusions}
In earlier attempts\cite{Slag1,Slag2,Slag3,Slag4}, we tried to build a 5-dimensional cosmic string without a warp factor and investigated
the causal structure. Here we considered on a warped 5D space time the "classical" self-gravitating Nielsen-Olesen vortex.
It seems possible that the absence of evidence of cosmic strings in observational data could be explained by our model, where the effective angle deficit resides in the bulk
and not in the brane.

In the super-massive case of Laguna and Garfinkle\cite{Lag2}, i.e., where the linear mass per unit length $G \mu >> 10^{-6}$, a continuous transition occurs from a conical space time to a Kasner-type with a curvature singularity at finite distance of the core , when the energy scale of the symmetry breaking increases.
Super-massive cosmic strings are however inconsistent with observation and would have to be formed before an  inflationary era.
In our induced brane space time we find a different result. On the brane, there is no conical space time measured far from the core of the string.
The solutions don't change significantly for increasing symmetry breaking scale $\eta$. We find an exact expression for the warp factor, which will warp down the found
Kasner-like solutions.

The next step in evaluation of this model, will be the dynamical behavior of the brane, by solving the full time-dependent equations.
This subject in under study by the authors.

\section*{Acknowledgments}

We are grateful to Professor Stanley Deser of Brandeis University at Waltham, for reading our manuscript.
\appendix
\section{Summary  of the U(1)-gauge cosmic string in 4D}
For the reader unfamiliar with cosmic strings, we will give a brief overview of the main features.
It is believed that topological defects, remnants of phase transitions caused by spontaneously broken symmetry, hold the promise of finding
a satisfactory model for the formation of the large scale structure of the galaxy distribution in our universe. Cosmic strings play a crucial role in these models. Apart from their possible astrophysical role, they are fascinating objects in their own right and can give rise to a rich variety of unusual physical phenomena such as the (2+1) dimensional spinning point source, which admits closed time like curves. See however\cite{Deser1,Deser2}. The book of A. Vilenkin and E. Shellard\cite{Vil} presents a fine complete overview of all the
features of the cosmic string.
The thickness of a cosmic string is $\eta^{-1}$, where $\eta$ defines the energy scale of symmetry breaking. For GUT-scales, $\eta\sim 10^{16}GeV$,
leading to a thickness of $\delta \sim 10^{-30} cm$. However, in analyzing the properties   of the string, one cannot neglect their thickness.
When solving the coupled Einstein-scalar-gauge field equations, one finds that the space time outside  the string exhibits an angle deficit, i.e.,
Minkowski minus a wedge and is of order $\Delta\theta \approx \kappa_4^2 \mu$ , where $\mu$ is the linear energy density of the string.The ratio
$\frac{\mu}{\eta^2}$ depends only on the ratio of the scalar- and gauge masses. The Lagrangian is\cite{Lag}
\begin{equation}
{\cal L}=\frac{1}{2\kappa_4^2}R-\frac{1}{2}D_\mu\Phi(D^\mu\Phi)^*-V(\mid\Phi\mid)-\frac{1}{4}F_{\mu\nu}F^{\mu\nu},\label{app1}
\end{equation}
where $D_\mu\equiv\nabla_\mu+ieA_\mu$, $F_{\mu\nu}=\nabla_\mu A_\nu -\nabla_\nu A_\mu$ and $ V(\Phi)=\frac{\beta}{8}(\mid\Phi\mid^2-\eta^2)^2$
The field equations then become
\begin{equation}
G_{\mu\nu}=k_4^2 T_{\mu\nu},\label{app2}
\end{equation}
\begin{equation}
D_\mu D^\mu \Phi -2\frac{\partial V}{\partial \Phi^*}=0,\label{app3}
\end{equation}
and
\begin{equation}
\nabla^\nu F_{\mu\nu}-\frac{1}{2}ie\Bigl[\Phi(D_\mu\Phi)^*-\Phi^*(D_\mu \Phi)\Bigr] = 0,\label{app4}
\end{equation}
with $T_{\mu\nu}$ the energy momentum tensor.
Further, one has $ m_\Phi ^2=\beta \eta^2$ and $ m_A^2=e^2\eta^2$, so $ \frac{m_A^2}{m_\Phi^2}=\frac{e^2}{\beta}\equiv\alpha$.
The scalar, gauge and gravitational fields take the form
\begin{eqnarray}
\Phi=Q(r)e^{i\varphi}\qquad A_\mu=\frac{P(r)-1}{e}\nabla_\mu\varphi,\label{app5}
\end{eqnarray}
\begin{equation}
ds^2=e^{A(r)}(-dt^2+dz^2)+dr^2+K(r)^2e^{-2A(r)}d\varphi^2.\label{app6}
\end{equation}
If one re-scales $Q\equiv\eta X,  r\rightarrow\frac{r}{\eta\sqrt{\beta}}$ and $ K\rightarrow\frac{K}{\eta\sqrt{\beta}}$, then the radii of the core false vacuum and magnetic field tube are $r_\Phi\approx 1, r_A^2\approx\frac{1}{\alpha}$ and one has only two free parameters $\alpha$ and
$\eta$. The set of equations become

\begin{equation}
\partial_{rr}K=\frac{1}{2}\kappa_4^2\eta^2\Bigl[-\frac{3}{4}K(X^2-1)^2-2e^{2A}\frac{X^2P^2}{K}+\frac{e^{2A}}{\alpha K}(\partial_rP)^2\Bigr],\label{app7}
\end{equation}
\begin{equation}
\partial_{rr}A+\frac{\partial_r A\partial_r K}{K}=\kappa_4^2\eta^2\Bigl[\frac{e^{2A}(\partial_r P)^2}{\alpha K^2}-\frac{1}{4}(X^2-1)^2\Bigr],\label{app8}
\end{equation}
\begin{equation}
\partial_{rr}X=-\frac{\partial_r K \partial_r X}{K}+\frac{1}{2}X(X^2-1)+\frac{XP^2e^{2A}}{K^2},\label{app9}
\end{equation}
\begin{equation}
\partial_{rr}P=-2\partial_r P \partial_r A+\frac{\partial_r P \partial_r K}{K}+\alpha X^2P.\label{app10}
\end{equation}
There don't exist a solution in closed form. A typical numerical solution is plotted in figures 6 and 7. We used a standard two-point boundary value
program with initial and boundary conditions:
\begin{eqnarray}
e^{A(0)}=1,\quad A_r(0),\quad K(0)=1, \quad X(0)=0, \quad P(0)=1,\cr  X(\infty)=1,\quad P(\infty)=0.\label{app11}
\end{eqnarray}

\begin{figure}
\centerline{
\includegraphics[width=4cm]{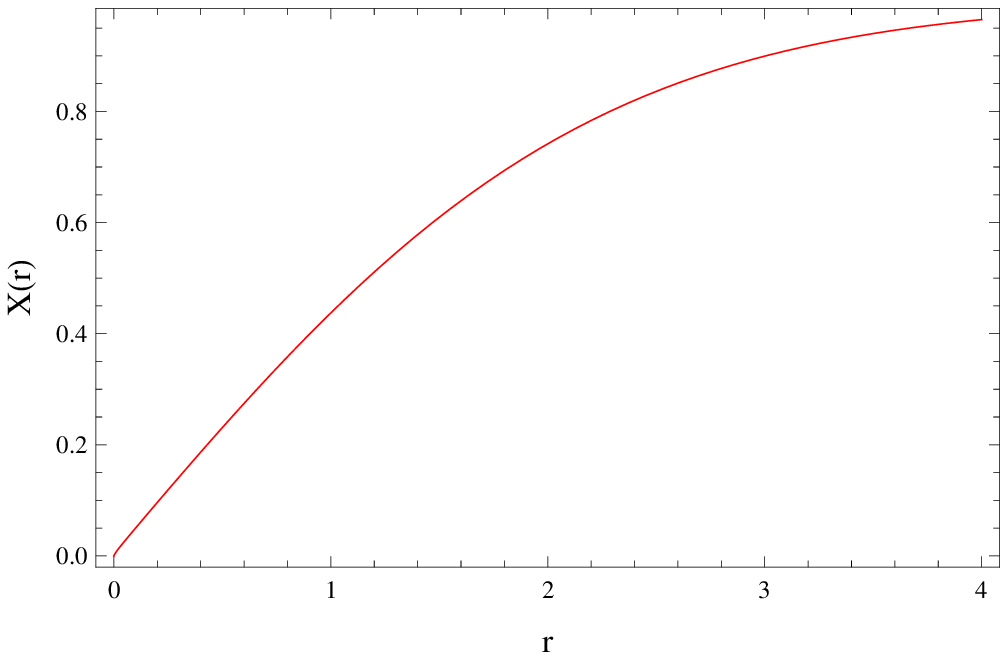}
\includegraphics[width=4cm]{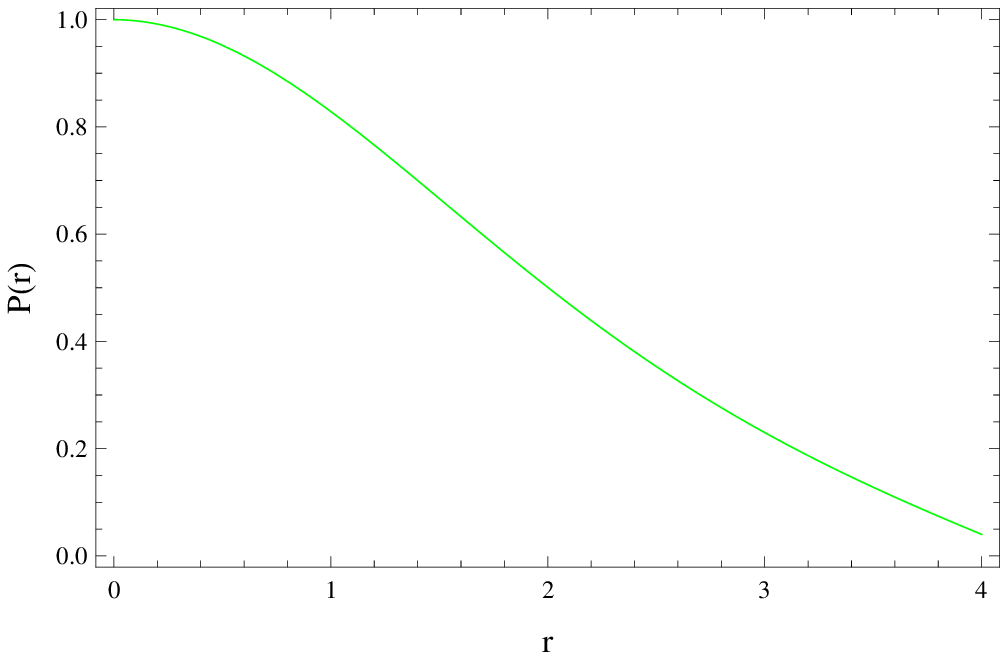}
\includegraphics[width=4cm]{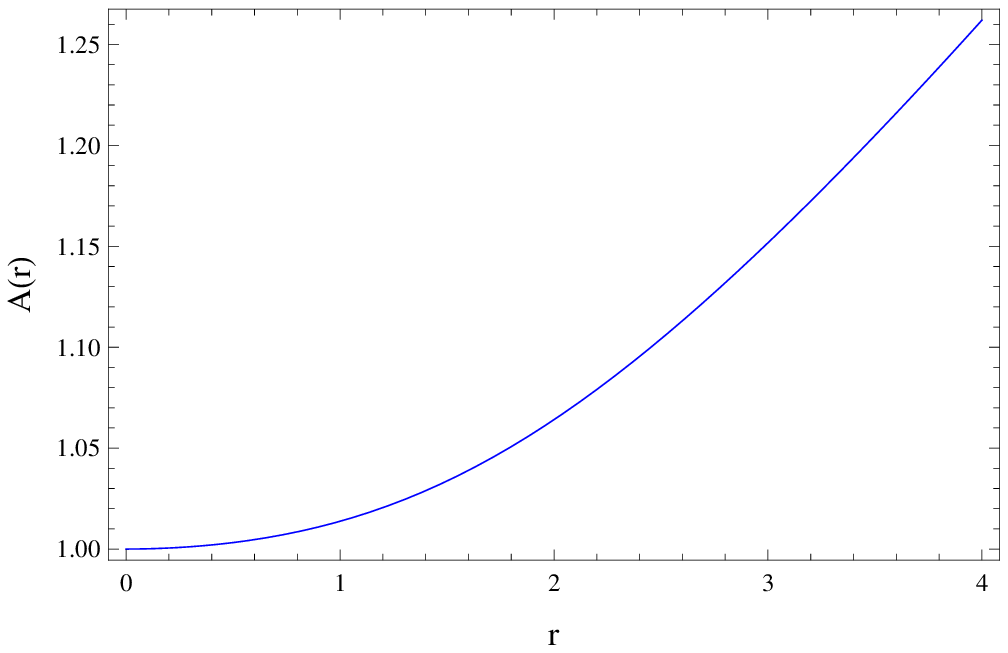}
\includegraphics[width=4cm]{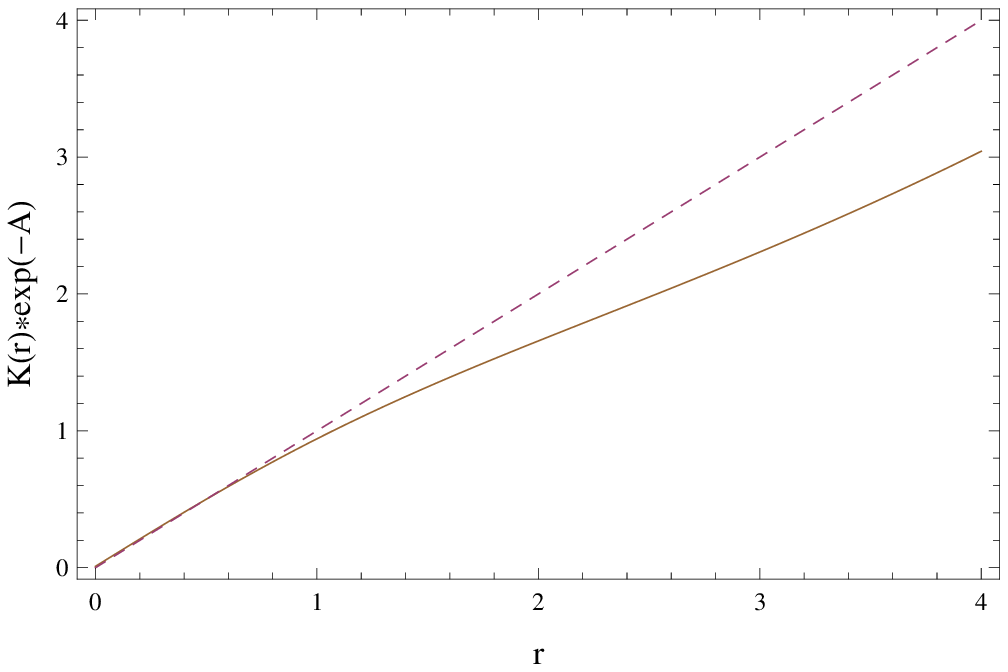}}
\vspace*{8pt}
\caption{Characteristic solution of $X,P,A$ and $e^{-A}K$ of the U(1) gauge string for $\eta=0.1, \alpha =0.25$ using a "shooting" method. The dashed line represents Minkowski space time.\label{f6}}
\end{figure}
\begin{figure}
\centerline{
\includegraphics[width=5cm]{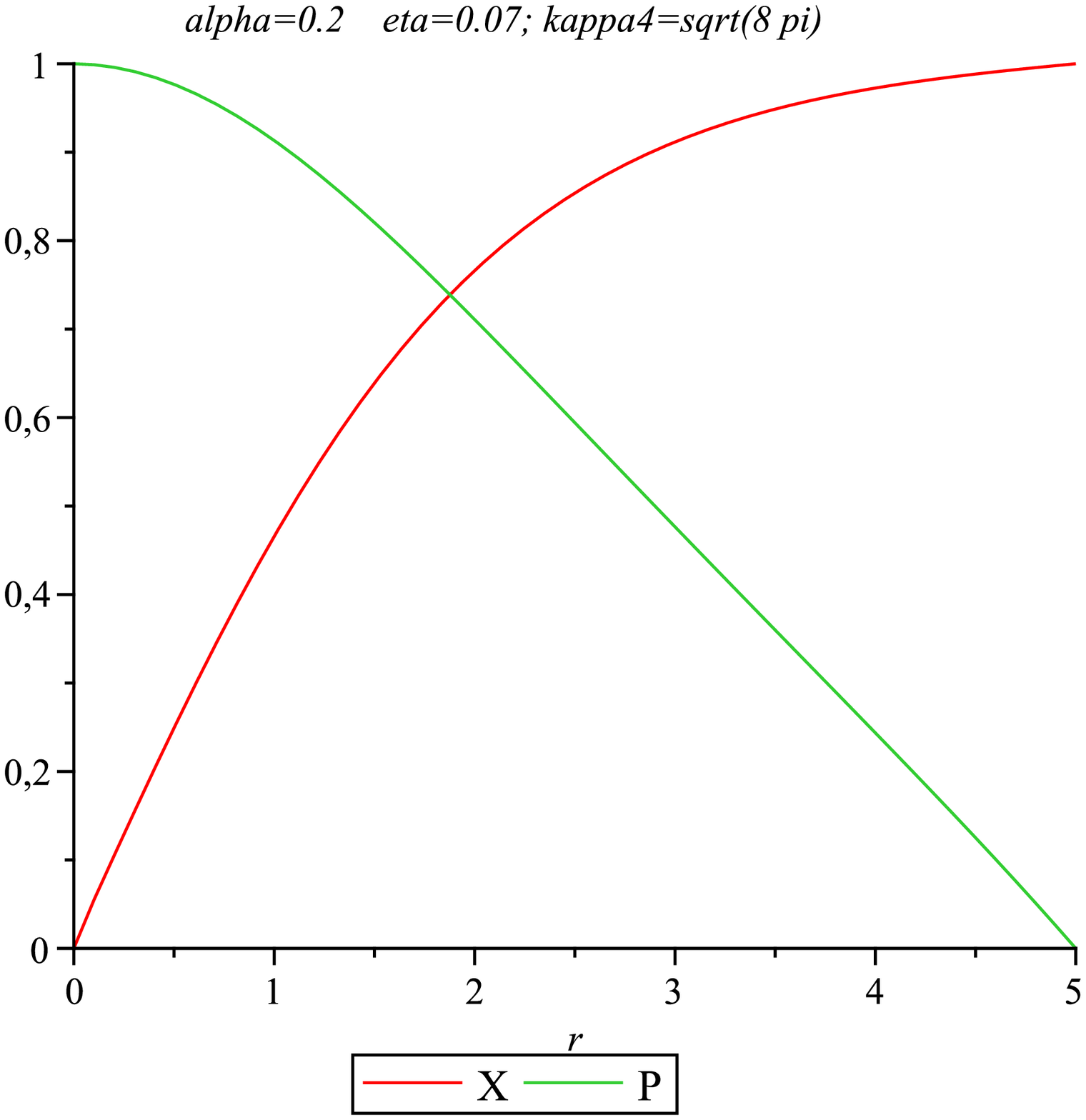}
\includegraphics[width=5cm]{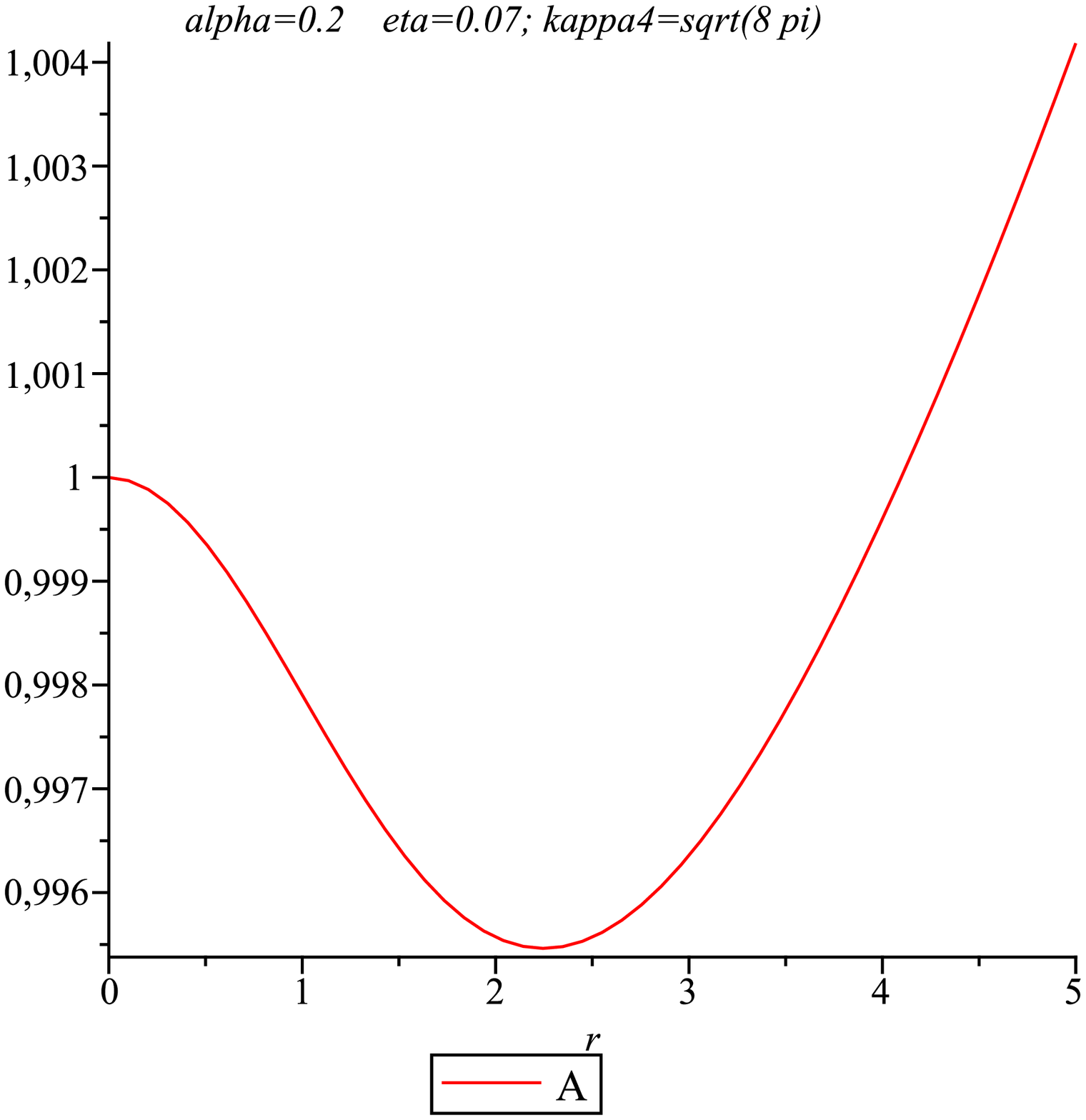}
\includegraphics[width=5cm]{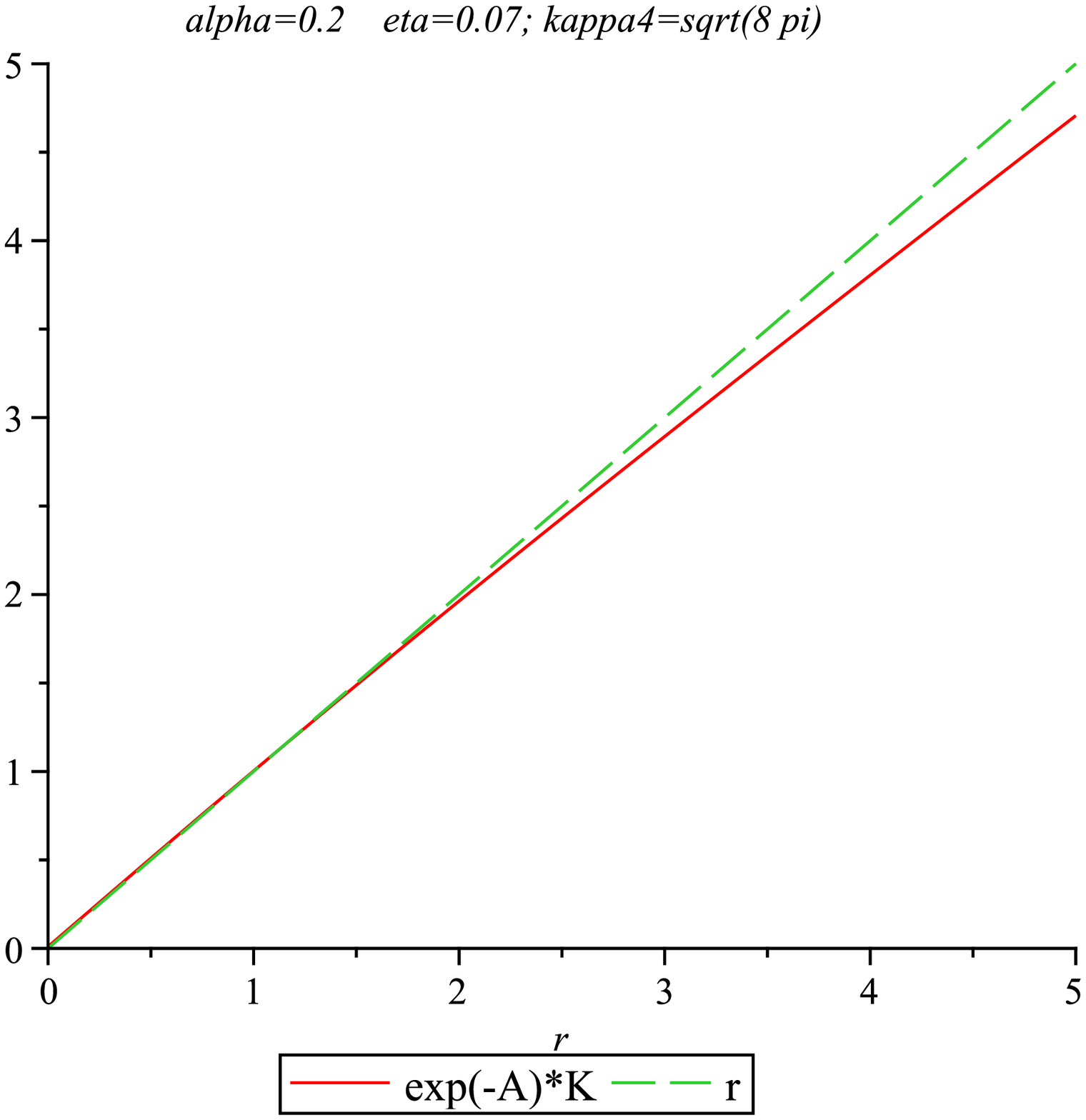}}
\vspace*{8pt}
\caption{Characteristic solution of the U(1) gauge string using a two point boundary routine for $\eta=0.07, \alpha =0.2$ and $\kappa_4=\sqrt{8\pi}$ as in the Laguna Castillo-Matzner solution. We see an angle deficit in the graph of $e^{-A}K$. The dashed line represents Minkowski space time.\label{f7}}
\end{figure}

Following Garfinkle\cite{Garf} one  defines
\begin{eqnarray}
\Theta_1\equiv K\partial_rA \qquad \Theta_2\equiv \partial_rK,\label{app12}
\end{eqnarray}
and the stress energy of the fields as
\begin{equation}
T_{\mu\nu}=\sigma \hat{k}_t\hat{k}_t+\varrho_z\hat{k}_z\hat{k}_z+\varrho_\varphi \hat{k}\varphi\hat{k}\varphi+\varrho_r\hat{k}_r\hat{k}_r,\label{app13}
\end{equation}
with $ \sigma =-\varrho_z $ and where  $\hat{k}_t = e^{\frac{A}{2}}(\frac{\partial}{\partial t}), \hat{k}_z = e^{\frac{A}{2}}(\frac{\partial}{\partial z}) ,
\hat{k}_\varphi= Ke^{-A}(\frac{\partial}{\partial \varphi}) ,\hat{k}_r = (\frac{\partial}{\partial r})$ are the set of orthonormal vectors.
The field equations Eq.(\ref{app7}) and (A8) can then be written as
\begin{eqnarray}
\partial_r\Theta_1=\kappa_4^2 K(\varrho_\varphi+\varrho_r) \qquad \partial_r\Theta_2 =\frac{1}{2}\kappa_4^2K(3\varrho_r+\varrho_\varphi-2\sigma).\label{app14}
\end{eqnarray}
From the conservation of stress energy we obtain
\begin{eqnarray}
\hat{k}_\mu\nabla_\nu T^{\nu\mu}=\nabla_\nu(T^{\nu\mu}\hat{k}_\mu)-T^{\nu\mu}\nabla_\nu\hat{k}_\mu \cr = \partial_r(K\varrho_r)-\varrho_\varphi\partial_r K
+(\sigma +\varrho_\varphi)K\partial_r A=0.\label{app15}
\end{eqnarray}
One can then write the total derivative of a combination of $\Theta_i$\cite{Garf}as
\begin{eqnarray}
\partial_r\Bigl[\Theta_1(\Theta_2-\frac{3}{4}\Theta_1)\Bigr]=\partial_r\Bigl[k_4^2K^2\varrho_r\Bigr],\label{app16}
\end{eqnarray}
where one uses Eq.(\ref{app15}) and the two field equations.
If we assume that $\int_0^\infty K\sigma dr$ converges and that
$$\lim_{r\rightarrow \infty}K^2\sigma=0 ,$$
which are fairly weak assumptions about the stress-energy and  assuming suitable boundary condition at the axis, we then  obtain from Eq.{\ref{app16})
\begin{equation}
k_1(k_2-\frac{3}{4}k_1)=0,\label{app17}
\end{equation}
where $k_1$ and $k_2$ are the asymptotic values of $\Theta_1$ and $\Theta_2$  respectively. One also uses the fact that $ \sigma >|\varrho_r|$.
The solution $k_2=\frac{3}{4}k_1$ represents a Kasner-like solution\cite{Vil}.
For $k_1=0$ we then obtain $\partial_r A=0, \partial_r K=k_2$. The asymptotic metric becomes
\begin{equation}
ds^2=-e^{a_0}(dt^2-dz^2)+dr^2+e^{-2a_0}(k_2r+a_2)^2d\varphi^2,\label{app18}
\end{equation}
where $a_0$ and $a_2$ are integration constants.
So the metric field $K$ can be approximated for large $r$ by $(k_2r+a_2)$, where $k_2$ will be determined by  the energy scale $\eta$ and
the ratio $\frac{m_A}{m_\Phi}=\sqrt{\alpha}$.
This metric can be brought to Minkowski form by the change of variables
\begin{eqnarray}
r'=r+\frac{a_2}{k_2},\quad \varphi '=e^{-a_0}k_2\varphi,\quad t'=e^{a_0/2}t, \quad z'=e^{a_0/2}z,\label{app19}
\end{eqnarray}
where now $\varphi '$ takes values $0\leq\varphi'\leq 2\pi e^{-a_0}k_2$. So we have a Minkowski metric minus a wedge with angle deficit
\begin{equation}
\Delta\theta=2\pi(1-e^{-a_0}k_2).\label{app20}
\end{equation}

The angle deficit can we written as\cite{Garf}
\begin{eqnarray}
\Delta\theta=2\pi\Bigl[1-\lim_{r\rightarrow\infty}\frac{d}{dr}(e^{-A}K)\Bigr] \cr
=\kappa_4^2\int_{0}^{\infty}2\pi e^{-A}K\sigma dr+\frac{\pi}{2}\int_{0}^{\infty}e^{-A}K\Bigl(\frac{dA}{dr}\Bigr)^2dr.\label{app21}
\end{eqnarray}
Or
\begin{equation}
\Delta\theta =\kappa_4^2\mu+\frac{\pi}{2}\int_0^\infty e^{-A}K\Bigl(\frac{dA}{dr}\Bigr)^2 dr,\label{app22}
\end{equation}
with $\mu \sim \eta^2 $ the linear energy density of the string
\begin{equation}
\mu\equiv2\pi\int_0^\infty e^{-A}K\sigma dr,\label{app23}
\end{equation}
and $\sigma$ the $(tt)$ component of the energy momentum tensor.
The angle deficit will increase with the energy scale of symmetry breaking. Further,
$\frac{\mu}{\mu_0}\approx (\frac{m_A}{m_\Phi})^{-0.424}$ \cite{Jac}, where $\mu_0$ is the energy scale for $m_A =m_\Phi$.

For GUT scale, $\eta\sim 10^{16}$ GeV, so the mass per unit length is $G \mu\sim 10^{-6}$. Numerical analysis of super massive cosmic
strings\cite{Lag2}, where $G\mu\gg 10^{-6}$, shows that the solution becomes singular at finite distance of the string or the angle deficit becomes greater than $2\pi$.


\end{document}